\documentclass{article}
\usepackage{graphicx} 
\usepackage[a4paper,margin=1in]{geometry}
\usepackage[
backend=biber,
style=ieee,
sorting=none,
giveninits=true,
firstinits=false,
maxnames=99  
]{biblatex}
\usepackage[hidelinks]{hyperref} 
\usepackage[backend=biber]{biblatex}
\title{ANNE Apnea Paper}
\author{Dominik Luszczynski}
\date{March 2025}
\usepackage{graphicx}
\usepackage{amsmath}
\usepackage{amsthm}
\usepackage{amssymb}
\usepackage{ragged2e}
\usepackage{pdfpages}
\usepackage{pgfplots}
\usepackage{indentfirst}
\usepackage{listings}
\usepackage{enumerate}
\pgfplotsset{compat=1.18}
\usepackage{graphicx}
\usepackage{tikz}
\usepackage{multirow}
\usepackage{placeins}
\usepackage{makecell}
\usepackage{algorithm}
\usepackage{algpseudocode}
\usepackage{algorithmicx}
\usepackage{float}
\usepackage{fontawesome5}
\usepackage[labelformat=simple]{caption}
\DeclareCaptionLabelFormat{customlabel}{\textbf{#1 #2}}
\justifying
\begin{document}

\noindent \rule{\textwidth}{3pt}

\vspace{2mm}

\begin{center}
    \noindent \textbf{\LARGE Sleep Apnea Detection on a Wireless Multimodal Wearable Device Without Oxygen Flow Using a Mamba-based Deep Learning Approach}
\end{center}

\noindent \rule{\textwidth}{1pt}
\\

\begin{center}
    \textbf{Dominik Luszczynski$^{\dagger, 1, 2}$ \quad Richard Fei Yin$^{\dagger, 1, 2}$ \quad Nicholas Afonin$^{\dagger, 1, 2}$\\
    \quad Andrew S. P. Lim$^{1a, 2}$}
\end{center}

\noindent $^{1}$ University of Toronto Department of Medicine $^{a}$, Computer Science$^{b}$ \hfill Toronto, ON, Canada \\
$^{2}$ Sunnybrook Research Institute Department of Medicine Neurology Div. \hfill Toronto, ON, Canada

\def\thefootnote{$\dagger$}\footnotetext{These authors contributed equally.}

\begin{center}
\textsuperscript{\faEnvelope[regular]} \texttt{\{dominik.luszczynski, r.yin, n.afonin\}@mail.utoronto.ca}\\
\texttt{andrew.lim@utoronto.ca}
\end{center}
\section*{Abstract}

Objectives: In-laboratory polysomnography (PSG) is the reference standard for the diagnosis of sleep disordered breathing, and characterization of disease physiology along multiple dimensions.   However, it is labour intensive, poorly scalable, and can perturb sleep.  Numerous devices for home sleep apnea testing (HSAT), along with tools for automated analysis, have been developed to address these limitations.  Although several such tools have shown good performance at establishing a diagnosis of SDB at the recording level, most have uncertain performance at the individual event level, including differentiation of event type (e.g. central vs. obstructive apneas), and determination of precise event timing.  Moreover, while disruption of sleep architecture is a key feature of sleep disordered breathing, not all HSAT devices are able to return epoch-by-epoch sleep staging.  Here we present and evaluate a deep-learning model for diagnosis and event-level characterization of sleep disordered breathing based on signals from the ANNE One, a non-intrusive dual-module wireless wearable system measuring chest electrocardiography, triaxial accelerometry, chest and finger temperature, and finger phototplethysmography.\\

Methods: We obtained concurrent PSG and wearable sensor recordings from 384 adults attending a tertiary care sleep laboratory.  Respiratory events in the PSG were manually annotated in accordance with AASM guidelines.  Wearable sensor and PSG recordings were automatically aligned based on the ECG signal, alignment confirmed by visual inspection, and PSG-derived respiratory event labels were used to train and evaluate a deep sequential neural network based on the Mamba architecture.\\

Results: In 57 recordings in our test set (mean age 56, mean AHI 10.8, 43.86\% female) the model-predicted AHI was highly correlated with that derived form the PSG labels (R=0.95, p=8.3e-30, men absolute error 2.83).  This performance did not vary with age or sex.  At a threshold of AHI$>$5, the model had a sensitivity of 0.96, specificity of 0.87, and kappa of 0.82, and at a threshold of AHI$>$15, the model had a sensitivity of 0.86, specificity of 0.98, and kappa of 0.85.  At the level of 30-sec epochs, the model had a sensitivity of 0.93 and specificity of 0.95, with a kappa of 0.68 regarding whether any given epoch contained a respiratory event.  Among 3,555 detected respiratory events, and contrasting central apneas from other events, the model correctly classified 77\% of central apneas and 95\% of RERA/Hyp/Obs.  Moreover, model predicted mean respiratory event duration was moderately correlated with PSG-derived event duration (R=0.53, p=4.9e-3).  \\

Conclusions: Applied to data from the ANNE One, a wireless wearable sensor system without sensors on the face or the head, a Mamba-based deep learning model can accurately predict AHI and identify SDB at clinically relevant thresholds, achieves good epoch- and event-level identification of individual respiratory events, and shows promise at physiological characterization of these events including event type (central vs. other) and event duration.

\section*{Statement of Significance}

Although polysomnography is the most reliable approach to detecting obstructive sleep apnea (OSA), the annotation process is both time-consuming and expensive. Machine learning and AI models trained on non-intrusive wearable devices, provide a quick and low-cost alternative to diagnose sleep apnea. We developed a Mamba-based apnea detection model on an old and diverse dataset of 384 patients collected at a tertiary care sleep clinic using the ANNE One wearable device. The Mamba-based model is able to achieve good performance for detecting apnea events, in addition to matching human specialists at computing recording-level AHI. Finally, this research establishes a comprehensive framework for evaluating OSA detection models, extending existing approaches in current literature.

\section*{Keywords}

Deep Learning, Sleep Apnea Detection, Wearable Devices

\newpage

\section{Introduction}
Obstructive sleep apnea (OSA) is a common sleep disorder in which breathing reduces momentarily (hypopnea) or completely stops (apnea), causing short, repeated arousals throughout the night as individuals struggle to breathe. OSA is believed to affect more than 936 million adults of ages 30-69 \cite{apnea_prevalence:2025}, is associated with conditions such as acute coronary syndrome and hypertension, and is estimated to double medical expenditures for healthcare systems due to its role in cardiovascular diseases \cite{apnea_prevalence:2025, apneahypertension:2008, economicApnea:2013}. With 80-90\% of cases remaining undiagnosed in the United States, a quick, low-cost and reliable diagnostic method is essential to expand access to OSA care \cite{anne_baseline:2022}.\\

The best and most reliable approach to OSA detection is a polysomnography (PSG), where various sensors are attached to a patient to record physiological data throughout a night of sleep. Recordings are typically divided into contiguous 30-second or 60-second windows which are annotated individually. An oronasal thermal sensor is used to identify apneas by a 90\% drop in airflow for $\geq 10$ seconds while a nasal pressure transducer detects hypopneas by a 30\% drop in airflow for $\geq 10$ seconds coupled with an arousal or a drop in blood oxygen saturation (SPO2) of 3\% \cite{polysomnography:2019, aasmStandards:2012}. The number of apneas or hypopneas per hour of sleep, called the Apnea-Hypopnea Index (AHI), is often compiled to gauge a patient's apnea severity. Under the American Academy of Sleep Medicine (AASM)'s conventions, AHI is grouped in four brackets: no apnea ($< 5$), mild apnea ($[5, 15)$), moderate apnea ($[15, 30)$), and severe apnea ($\geq 30$) \cite{aasmStandards:2012}. \\

Although PSGs are powerful, the data collection and annotation process is expensive and time-consuming. Being conventionally performed in sleep clinics, PSGs may experience long wait times, pose challenges in accessibility for elderly, rural, or sick populations, and exhibit a "first-night effect", a reduction in sleep quality caused by the unfamiliar laboratory environment \cite{anne_baseline:2022}. Efforts to conduct at-home PSGs are furthermore complicated by the transportation and set-up of equipment. Consequently, wearable devices have gained ground as alternatives to PSGs that capture a smaller selection of signals, making them simpler and cheaper but less informative in their ability to distinguish apnea-related arousals from noise or respiratory effort-related arousals (RERAs). Similarly, automated methods to speed up PSG annotation have been explored, with deep learning methods such as vanilla neural networks, deep belief networks (DBNs), convolutional neural networks (CNNs), long short-term memory networks (LSTMs), or a combination of CNNs and LSTMs, exceeding 90\% accuracy for detecting apneas on a per time window basis \cite{automated_apnea_detection:2023}. \\

\vspace{-7mm}
\begin{center}
\begin{figure}[H]
    \centerline{\includegraphics[width=0.75\textwidth]{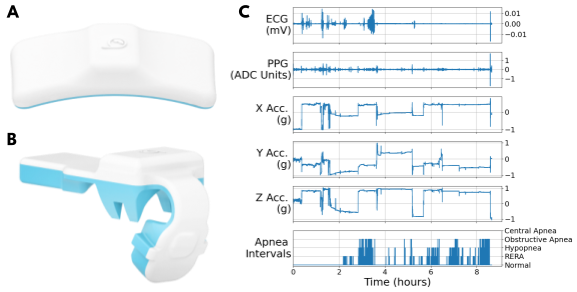}}
    \vspace{-3mm}
    \caption{The ANNE One's chest module (A), attached via an adhesive; the limb module (B), attached around a finger; and example raw signals with PSG-labelled apneas (C). Images by Sibel Health \cite{Anne-One-Limb, Anne-One-Chest}.}
    \label{fig:apnea-one}
\end{figure}
\end{center}
\vspace{-10mm}

This study combines wearable technologies and automated OSA detection by applying the Mamba architecture to OSA detection using sensor data from the ANNE One wearable device (Figure \ref{fig:apnea-one}):

\begin{itemize}
\item ANNE One is a pair of clinical-grade wireless devices mounted to the chest, where temperature, electrocardiography (ECG), and triaxial accelerometry (X/Y/Z channels) are measured; and the finger, where it measures temperature and photoplethysmography (PPG) \cite{anne_baseline:2022}. Although the system is low-cost and non-intrusive, the absence of an airflow sensor forces detection methods to infer rather than measure apneas, while the lack of an electroencephalography (EEG) signal makes sleep-wake disambiguation more difficult. In spite of this, annotated ANNE One labels have achieved accuracies of 90s in binary AHI severity classification (using PSG-annotated labels as ground truths) solely by analyzing pulse arrival time (PAT) derived from ECG and PPG signals \cite{anne_apnea:2022}.

\item Mamba is a selective state-space model (SSM) architecture for processing sequential data (e.g. channels of biosignals). It is competitive with the state-of-the-art transformer architecture on language and audio tasks while scaling linearly in running-time with sequence length, allowing it to process long sequences quicker than the quadratically-scaling transformer \cite{mamba:2023}. The Mamba architecture (details in \cite{zhang:2025}, S.1) combines the SSM's ability to capture long-term dependencies with a selection mechanism that focuses on salient features in the sequence, making it suitable in theory for finding apnea clusters,  arousals, and changes to SPO2 or heart rate. Mamba has seen success in OSA detection, reaching 90s in segment-wise apnea classification accuracy on the Apnea-ECG dataset \cite{mamba_apnea:2025}, a single-channel dataset of ECG signals from PSGs \cite{apnea_ecg:2000}.
\end{itemize}

Research on automated OSA detection with wearable devices typically focuses on a single biomedical marker such as SPO2, ECG, EEG, and electrooculography (EOG) or a relatively small combination of signals \cite{ramachandran:2021, abdalrazaq:2024, sanchez:2025}. These studies are also commonly trained on high-quality signals on relatively younger (average age of 47.3 across datasets in \cite{abdalrazaq:2024}) and smaller datasets (63\% of datasets in \cite{abdalrazaq:2024} have $n < 100$), boasting accuracies of 80s-90s in binary apnea severity classification but raising concerns of generalizability to older and sicker populations, whose biosignals often exhibit anomalies and poorer signal quality.\\

This paper contributes to the pre-existing body of research by training on a old (average age of 56.57) and diverse dataset with disorders such as periodic limb movement syndrome, using a methodology that simulates poor signal quality characteristic of a technician-free environment. Additionally, we exploit ANNE One's rich combination of signals that theoretically allow for robust predictions even in periods of poor quality in some signals. Finally, we expand Mamba's viability to wearable-based OSA detection and study the architecture's resilience to noise.

\section{Method}

\vspace{-5mm}
\begin{center}
\begin{figure}[H]
    \centerline{\includegraphics[width=\textwidth]{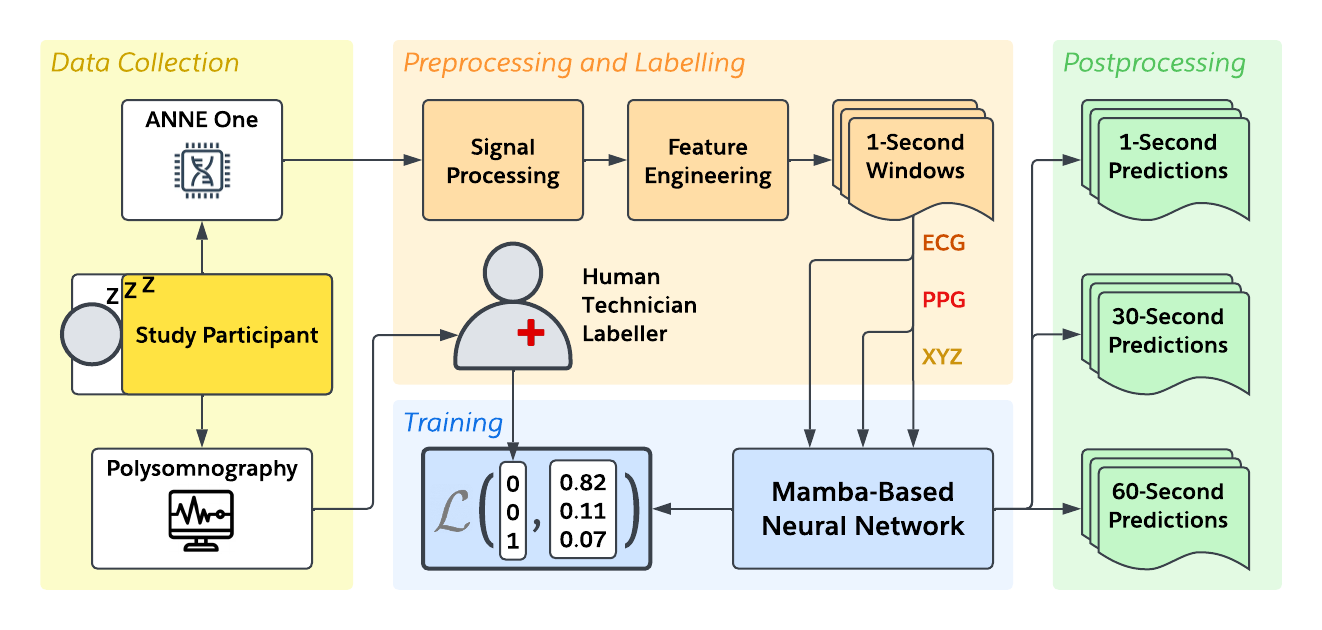}}
    \vspace{-5mm}
    \label{fig:apnea-pipeline}
    \caption{Overview of the study pipeline.}
\end{figure}
\end{center}
\vspace{-15mm}

\subsection{Dataset and Preprocessing}

The dataset originated from the same experiment as \cite{zhang:2025}, which was conducted at a tertiary care sleep clinic in Sunnybrook Health Sciences Centre (Toronto, Canada). Using ANNE One, overnight recordings were collected from 427 consenting participants (clinical characteristics in Table \ref{tab:demographics_table}) who then underwent a simultaneous PSG using a Grael PSG system (Compumedics, Victoria, Australia). A cross correlation analysis was performed to remove heavily misaligned recordings. For each recording, the number of valid matching points were computed for seconds of lag ranging -100 to 100 seconds, along with the median number of matching points as a baseline. The cross correlation score was computed by subtracting the baseline from the maximum number of matching points, and scores below 4\% of the maximum were not considered. After removing 39 patients with poor cross-correlation alignment and 4 patients who did not sleep at all, the final dataset contained 384 patients and 3159.87 hours of sleep, an increase from \cite{zhang:2025} as more data had been collected thereafter. In the interest of simulating a home environment, ANNE signals were not monitored in real-time. Additionally, the portion of ANNE recordings captured prior to the PSG recording was labeled as wake based on direct observation, resulting in a preponderance of awake time. \\

\begin{table}[H]
    \caption{Demographic characteristics of study participants for each dataset split, presented as mean $\pm$ standard deviation.}
    \label{tab:demographics_table}
    \small
    \vspace{-2mm}
    \begin{tabular}{l|ccc}
        \hline
        \textit{Characteristic\phantom{\Large{A}}} & \textit{Training Set} & \textit{Validation Set} & \textit{Test Set} \\
        \hline
        Sample Size (\# female)\phantom{\Large{A}} & 289 (129) & 38 (15) & 57 (25) \\
        Age (years) & 56.15 $\pm$ 18.46 & 59.10 $\pm$ 13.94 & 56.24 $\pm$ 15.96 \\
        Apnea-Hypopnea Index (AHI, events/hour) & 10.42 $\pm$ 15.64 & 13.58 $\pm$ 22.45 & 10.75 $\pm$ 16.05 \\
        Body Mass Index (BMI, kg/$\text{m}^2$) & 27.96 $\pm$ 5.94 & 28.26 $\pm$ 6.24 & 26.87 $\pm$ 4.82 \\
        Periodic Limb Movement Syndrome (PLMS, events/hour) & 12.45 $\pm$ 22.94 & 18.66 $\pm$ 31.02 & 17.84 $\pm$ 33.29 \\
        Total Sleep Time (hours) & 5.07 $\pm$ 1.53 & 5.07 $\pm$ 1.55 & 5.14 $\pm$ 1.21 \\
        Sleep Efficiency (\%) & 71.60 $\pm$ 18.96 & 71.91 $\pm$ 17.13 & 72.29 $\pm$ 15.66\\
        \hline
    \end{tabular}
\end{table}

Recordings were manually annotated by four technicians based on PSG signals. A binary apnea classification scheme was adopted by grouping all apnea events (hypopneas, central apneas, and obstructive apneas) into an apnea label with the exception of RERAs, which were collapsed into non-apneas because RERAs are differentiated with a signal not provided from ANNE One.\\


The data was preprocessed similarly to \cite{zhang:2025}. The ECG, PPG and XYZ signals were recorded at 512 Hz, 128 Hz and 210 Hz respectively. To remove their baseline, the signals were subtracted by a copy that was high-pass filtered using a second-order Butterworth filter with a 0.05 Hz cutoff. A Short-Time Fourier Transform-based spectrogram analysis was performed on the XYZ signals, reducing the sample rate to 1 Hz. Every fifth frequency bin was selected, with others discarded, to reduce feature count and avoid overfitting. The frequency features were then log-transformed. \\

Various features were extracted from the ECG, PPG, and XYZ signals. The \textit{pulse arrival time} (PAT) is defined as the delay from the R-peak of the ECG to the arrival of the pulse wave in the PPG \cite{zhang:2025}. The computation of the PAT involved matching the ECG peaks and PPG feet from the ANNE recordings by considering all possible lags from 0 to 1000 msec, and selecting the value that maximizes the alignment between ECG peaks and PPG feet. Then, for every pair of ECG R-peaks and PPG-foots, the PAT of the heartbeat was computed. Additionally, PPG was used to calculate SPO2 and SPO2 desaturation, while body roll ($\phi$\footnotemark) and pitch ($\theta$\footnotemark) were calculated from XYZ. Three ECG-based features were derived: the absolute ECG value, a rolling standard deviation over a 5-second window, and the ECG signal's power defined as the sum of the frequency bins from 0.1 to 0.5 Hz. The same three features were extracted from the PAT signal with the baseline subtracted. \\

\def\thefootnote{$\dagger$}
\footnotetext{ \normalsize  $\phi = \arccos\Big(\frac{z}{\sqrt{x^2 + y^2 + z^2}}\Big)$}
\def\thefootnote{$\ddagger$}
\footnotetext{ \normalsize  $\theta = \text{sign}(y)\arccos\Big(\frac{x}{\sqrt{x^2 + y^2}}\Big)$}




\subsection{Model}

\begin{figure}[H]
    \centerline{
    \includegraphics[width=1.15\textwidth]{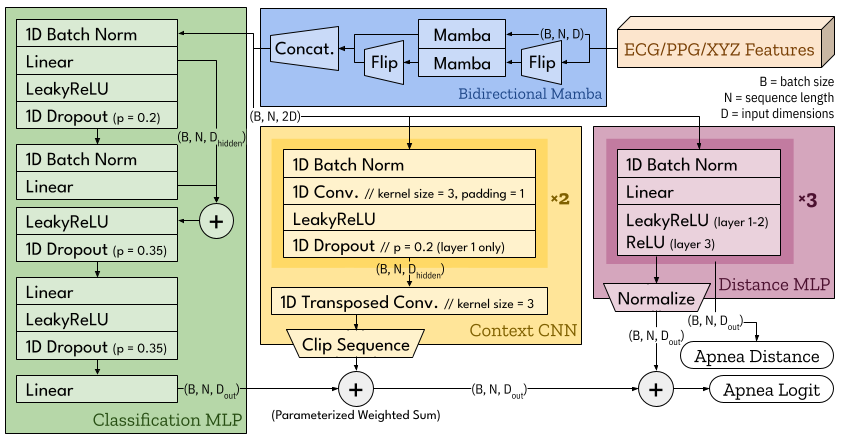}
    }
    \caption{Overview of model architecture, with intermediate tensor shapes specified in parentheses. Hyperparameters were determined experimentally (see Section \ref{model-params} in the Appendix for a comprehensive print of model parameter values). The apnea logits (bottom-right) are used for training, while the unnormalized apnea distances are used for postprocessing. For binary apnea classification, $D_{\text{out}} = 1$.}
    \label{fig:model-architecture}
\end{figure}

Inspired by the YOLO (You Only Look Once) architecture for multi-class object detection, where the model predicts class probabilities and bounding boxes for predicted classes \cite{yolo:2016}, the proposed model contains a classification and regression channel: the former predicts the probability of a time window containing an apnea, and the latter predicts the distance to the closest non-apnea measured in the number of windows (equal to 0 for non-apnea windows). The model sends the derived features into a 4-layer bidirectional Mamba block, whose output is sent to three blocks, as seen in Figure \ref{fig:model-architecture}:

\begin{enumerate}
    \item \textbf{Classification MLP}: A 3-layer MLP with skip connections that computes the logit probability of a 1-second time window containing an apnea. Batch normalization is omitted in the last layer to allow for separation of logits deeper in the network.
    \item \textbf{Context CNN}: Although Mamba is exceptional at learning long-term patterns, in instances such as computer vision tasks, Mamba struggles to outperform CNNs due to its poor capturing of short-term dependencies \cite{mamba_local:2024, mamba_apnea:2025}. Thus, a 2-layer CNN is incorporated to capture local patterns. A transposed 1D convolution is finally applied to allow predictions at time windows to affect neighbouring windows. Clipping is performed at the sequence's start and end to keep the sequence length constant.
    \item \textbf{Distance MLP}: A 3-layer MLP outputs the predicted distance to the closest non-apnea time window (in \# of windows). The last activation is substituted with ReLU to constrain distances to non-negative values.
\end{enumerate}

The outputs of the classification MLP and context CNN are multiplied by a parameterized weight and summed. To incorporate apnea distance information into the apnea final logit, the output of the distance MLP, normalized across the sequence length axis, is summed to the result. The normalization function is applied element-wise:

$$f(x) = \begin{cases}\frac{x}{2(x + 1)}, & \text{if } x > 0\\-0.5, & \text{if } x = 0\end{cases}$$

\subsection{Postprocessing}
\label{sec:postprocessing}

At inference time, Monte Carlo (MC) Dropout \cite{mcDropout:2015} was used to obtain a more stable estimate of the predicted logits by running 20 forward passes through the neural network with dropout terms active and averaging the results. The logits were then adjusted based on predicted unnormalized distances as described in Algorithm \ref{alg:postprocessing} to remove false positives and break up unfeasibly long intervals of apnea. \\

Predicted AHI for a recording was calculated by counting the number of predicted apnea intervals, then dividing by the total sleep time (TST) in hours. The sleep-wake model developed in \cite{zhang:2025} for the ANNE One device was used to identify intervals of sleep and calculate the TST for each recording. Note that, Figure \ref{fig:sleep-scatter-blandaltman} in the Appendix confirms that using the sleep-wake model compared to the ground truth sleep labels, does not degrade AHI correlation.

\begin{algorithm}
    \textbf{Input:} Recording-wise apnea logits $\textbf{l} \in \mathbb{R}^n$, unnormalized distances $\textbf{d} \in \mathbb{R}_{\geq0}^n$, predictions $\textbf{p} \in \{0, 1\}^n$, and the logit threshold $t \in \mathbb{R}$
    \begin{algorithmic}[1]
    \caption{Postprocessing algorithm for apnea classification.}
    \label{alg:postprocessing}
    \State $\textit{Remove apneas that are less than 10 seconds}$
    \If{$\max(\textbf{d}) < 3$} \Comment{Filter out false positives based on distances}
        \For{interval \textbf{in} $ \text{FindApneaIntervals}
        (\textbf{p})$}
            \State $\textit{Decrement } \textbf{l} \textit{ by } 1 \textit{ from } \text{interval start} \textit{ to } \text{interval end}$
        \EndFor
        \State $\textbf{p} \gets \text{Threshold } \textbf{l} \text{ by } t$
    \EndIf
    \For{interval \textbf{in} $\text{FindApneaIntervals}(\textbf{p})$} \Comment{Break up long intervals of apnea}
        \If{interval length $> 90$} 
            \State \textit{Decrement} \textbf{l} \textit{by} 5 \textit{for positions i from} interval start \textit{to} interval end \textit{where} $d_i \leq 0.5$
        \EndIf
    \EndFor
    \State $\textbf{return } \text{Threshold } \textbf{l} \text{ by } t$
    \end{algorithmic}
\end{algorithm}

\FloatBarrier

To generate predictions for $n$-second windows, all 1-second labels and predictions were grouped in contiguous buckets of size $n$. The prediction or label for a $n$-second time window was defined as the predicted presence of apnea in any of the inner 1-second time windows.



\subsection{Training}

Due to the severe class imbalance ($>90\%$ non-apneas), recordings were assorted into buckets of nearly equal size according to the apnea severity (AHI of $0, \leq 0.5, \leq 1, \leq 2.5, \leq 5, \leq 10, \leq 15, \leq 20, \leq 30, >30$), then a random stratified split was performed to determine the training, validation, and testing sets making up 75\%, 10\% and 15\% of the dataset respectively (see Table \ref{tab:demographics_table}). Hold-out validation was performed with the testing set removed, where five models were trained on different random stratified splits to determine the training and validation sets from the remaining recordings. The model with the best validation performance, defined as the highest AHI correlation with the ground truth combined with the highest balanced accuracy, was chosen was the final model for evaluation.\\

Training was performed with an Adam optimizer and learning rate of 0.001 with a weight decay of $10^{-5}$. To avoid overfitting to recordings, the model was trained on random 5000-second subsets of a recording at a time with a batch size of 4. The final model was chosen based on what maximized the sum of the intersection over union (IOU) and area under the precision-recall curve (AUPRC) for the validation set. \\

The loss function has three components which reflect the model's multi-channel output. The classification channel uses a combination of weighted cross-entropy and dice loss \cite{diceloss:2017}, defined as

\vspace{-4mm}
\begin{align*}
    \mathcal{L}_{\text{CE}} &= -w_i[y_i \log(\sigma(l_i)) + (1 - y_i) \log(1-\sigma(l_i))]\\
     \mathcal{L}_{\text{DICE}} &= 1 - 2 \times \frac{\sigma(l_P)y + \epsilon}{\sigma(l_P) + y + \epsilon} \quad \text{\cite{diceloss:2017}}
\end{align*}

\noindent for time window $i \in \mathbb{N}$, logit $l_i \in \mathbb{R}$, label $y_i \in \{0, 1\}$, sigmoid function $\sigma(x)=\frac{1}{1 + e^{-x}}$, class weight $w_i$, and $\epsilon = 10^{-6}$ which prevents division by zero. Akin to the YOLO regression loss, the distance loss is

\vspace{-4mm}
\begin{align*}
    & \mathcal{L}_{\text{DIST}} = (\sqrt{d_P} - \sqrt{d_T})^2 \quad \text{\cite{yolo:2016}}
\end{align*}

\noindent for predicted distance $d_P$ and ground truth distance $d_T$, where the square root ensures that the magnitude of error scales with distance itself. Finally, a mask $m$ (equal to 1/0 if a time window is currently sleep/wake) was applied to ensure that only sleep times were factored into the loss, preventing the prevalent awake time from affecting the quality of non-apnea predictions. The final loss is defined as

\vspace{-4mm}
\begin{align*}
    & \mathcal{L} = m \times (\mathcal{L}_{\text{CE}} + \alpha \mathcal{L}_{\text{DICE}} + \mathcal{L}_{\text{DIST}})
\end{align*}

\noindent where $\alpha = 1.5$ balances the magnitudes of the loss components. Loss was calculated per batch and time window and averaged over both axes. \\


\subsection{Statistical Analysis}
\label{sec:statistical_analysis}

Given the rapid emergence of automated apnea detection systems, the industry lacks a "standard" method of evaluation for such models. As such, a variety of statistical analyses and methods of evaluation were applied to the model's outputs, providing a thorough overview of performance, as well as the model's strengths and weaknesses. Some are more traditional, such as AHI correlation, while others, such as event-wise metrics, are intended to prompt more in-depth analysis of apnea detection models, and the furthering of evaluation standards. \\

First, performance of the model was evaluated on a \textit{recording-wise} basis for each test set recording by computing a patient's predicted AHI and comparing to the ground-truth AHI through linear regression and Bland-Altman plots. Apnea severity was analyzed by discretizing AHI into four classes using the thresholds 5/15/30 and computing the balanced accuracy, weighted (w.) precision, w. recall, specificity, w. F1, Matthew's Correlation Coefficient (MCC) and Cohen's Kappa ($\kappa$). Balanced accuracy and weighted metrics were chosen to account for the substantial class imbalance. In addition, a new metric TAO (True Apnea Overlap) which measures the percentage of true apnea intervals which are covered by a predicted apnea is used to determine how well the model classifies apnea intervals. \\


A similar methodology was used for \textit{apnea duration} analysis, where the total and average apnea durations in each test recording were calculated and plotted against the predicted total/average apnea durations in a linear regression plot and Bland-Altman plot. \\

The model was further analyzed on a \textit{segment-wise} level by aggregating all test set time windows spent asleep, then computing the same classification metrics for apneas and non-apneas. To allow for direct comparison to related literature, metrics were also computed for predictions in a 30-second and 60-second window resolution. \\

Two ablation studies were conducted -- an \textit{architectural} and \textit{feature} ablation study -- to determine the importance of each component and feature set in the proposed model. In the architectural ablation study, the Context CNN and Distance MLP were removed and the bidirectional Mamba block was replaced with a bidirectional LSTM. In the feature ablation study, the original model was trained on various subsets of signals from the chest and limb modules. \\

Owing to the significant presence of noise in our dataset, the model was assessed for robustness to poor signal quality. Segment-wise metrics were computed on time windows that were grouped by signal quality level, while linear regression was conducted between a recording's mean average error (MAE) in AHI and the patient's age, sex and PLMS. \\

Finally, we propose an \textit{event-wise }method of evaluation that treats the apnea detection task as one of event detection. This approach draws on methods used in analogous tasks, like seizure detection or arrhythmia detection, and aims to quantify what fraction of events the model caught (true positives), missed (false negatives), or falsely detected (false positives). Note that we do not quantify true negatives in event-wise evaluation like we do for segment-wise evaluation. This is because when viewing each apnea as an event and not as a class, it is both irrelevant and infeasible to count how many "non-events" there are. We define the remaining metrics in terms of "sufficient overlap" of events, measured by the Jaccard Index (IoU), or what we will call the IoU threshold. We define a true positive (TP) event when a pair of true and predicted events share some minimum IoU; we define a false positive (FP) event as a prediction with no sufficiently overlapping true event; and we define a false negative (FN) event as a true event with no sufficiently overlapping prediction. We note a relationship to the TAO metric. Specifically, event-wise recall with an IoU threshold of 0 is equivalent to TAO, and measures what percentage of true apnea events have some non-zero overlap with a prediction. Accurately characterizing overlap between true events and predictions with both event-wise metrics and TAO enables a more in-depth analysis. \\

\begin{figure}[H]
    \centering
    \includegraphics[scale=0.45]{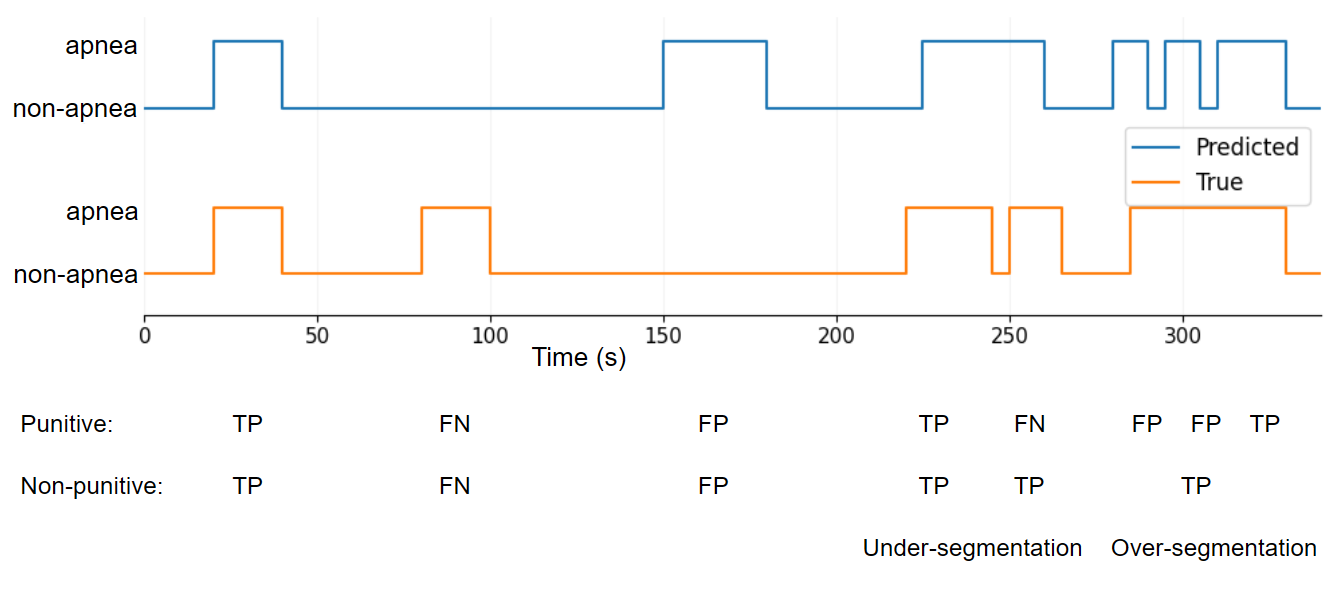}
    \vspace{-2mm}
    \caption{An example application of event-wise metrics.}
    \label{fig:eventwise_example}
\end{figure}

Two metrics help diagnose \textit{event-wise} performance: the over-segmentation index (OSI), and under-segmentation index (USI). Occasionally a model splits one event into multiple, or merges multiple events into one. OSI is defined as the number of extra predicted fragments divided by the total number of true events, and USI is the number of true events merged, divided by the total number of true events. For example, in Figure \ref{fig:eventwise_example}, the OSI would be 40\%, as two additional fragment are produced out of the five total true events, and the USI would be 20\%, as one true event is merged into another, out of the five total true events. To account for over- and under-segmentation, we distinguish \textit{punitive} (one-to-one) and \textit{non-punitive} (many-to-one) event-wise metrics. The former enforces a one-to-one match between each true event and predicted events, counting extra predicted fragments as false positives and unmatched true events as false negatives; the latter, by contrast, does not penalize fragmentation or merging, as long as each true event is sufficiently overlapped by at least one prediction. Both versions are illustrated in Figure \ref{fig:eventwise_example}. Further discussion and details relating to event-wise metrics are provided in Appendix \ref{sec:appendix:eventwise_metrics}.

\section{Results}
\subsection{Recording-Wise Performance}
\begin{figure}[H]
    \centering
    \includegraphics[scale=0.6]{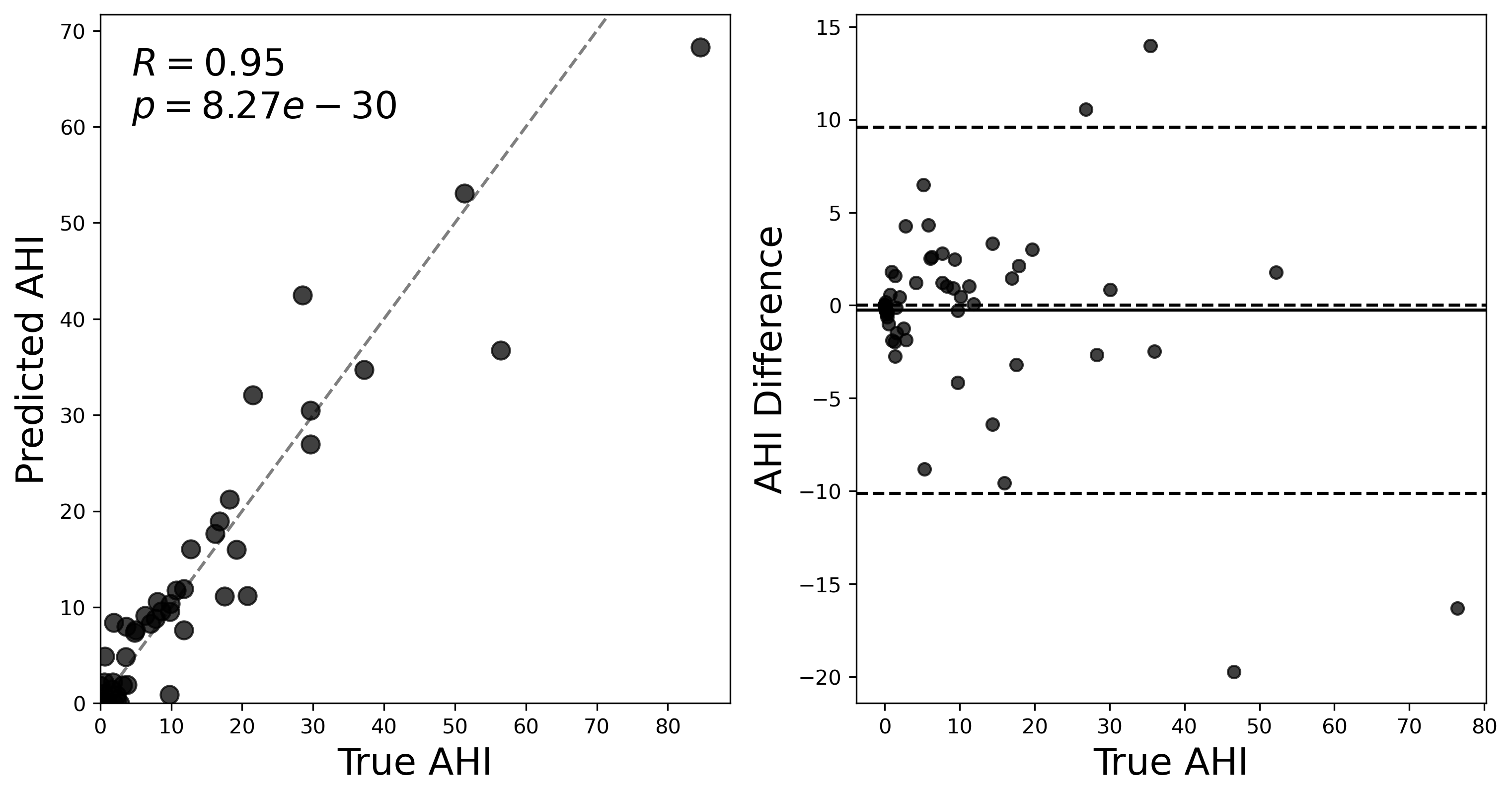}
    \vspace{-2mm}
    \caption{Scatterplot (left) and Bland-Altman plot (right) comparing the predicted and ground truth AHI across the test set. For the Bland-Altman plot, the solid line is the bias and the dotted lines in the Bland-Altman plot represent 0 as well as 1.96 standard deviations above and below the mean.}
    \label{fig:scatter-blandaltman}
\end{figure}

AHI values computed from model predictions are compared with labeled AHI values from the test set in Figure \ref{fig:scatter-blandaltman}, where predicted AHI displays a very strong positive correlation with true AHI (left, $R=0.95$, $p < 0.001$, Mean Absolute Error (MAE) $= 2.83$). The Bland-Altman plot (right) suggests that although the variance in residuals increases with actual AHI, the predictions themselves are unbiased (bias = $-0.28$).

\begin{figure}[H]
    \centering
    \includegraphics[scale=0.8]{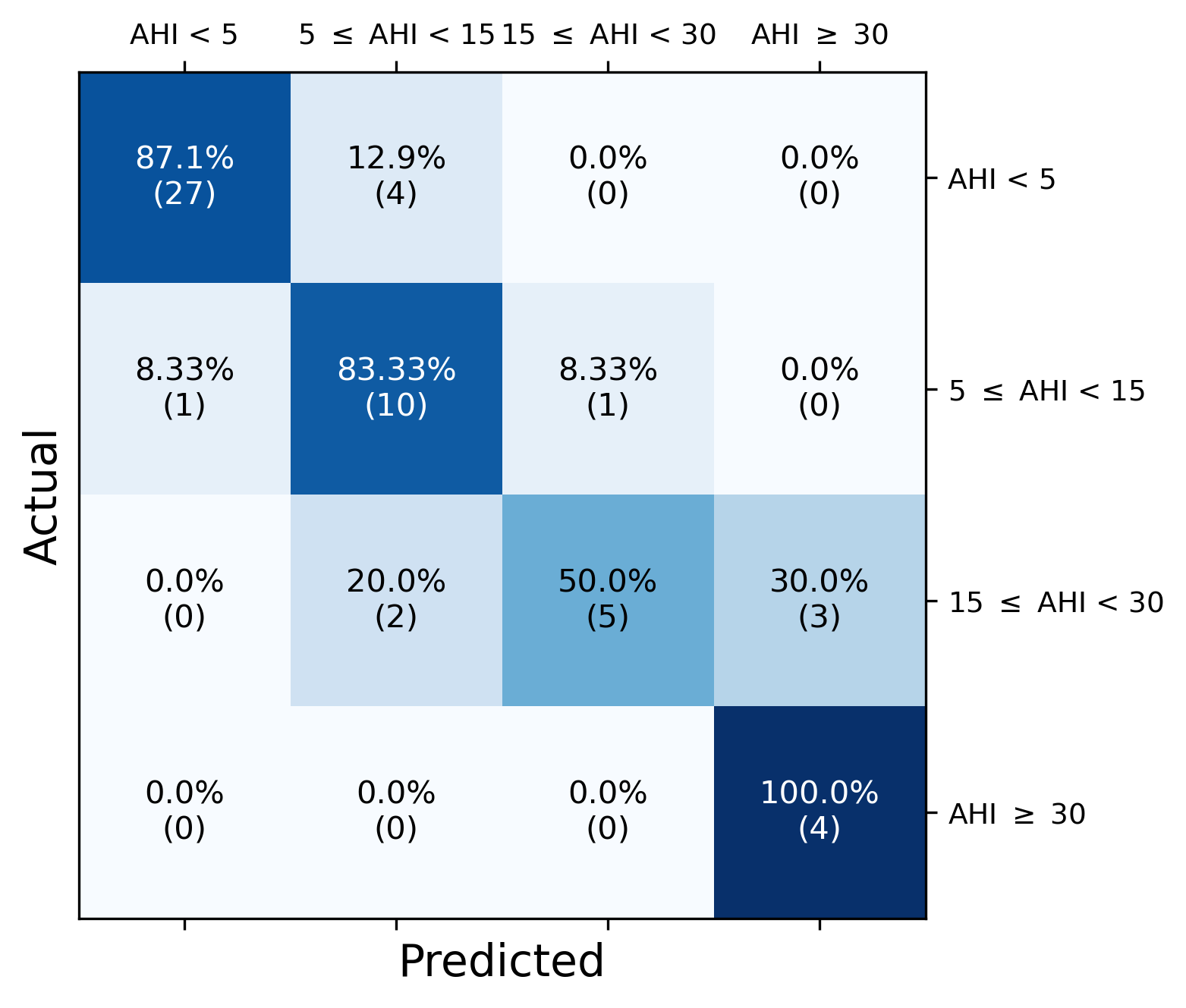}
    \vspace{-4mm}
    \caption{Confusion matrix of recording-level apnea severity on the test set.}
    \label{fig:confmat-ahitype}
\end{figure}

The model accurately distinguishes classes in a 4-class apnea severity setting as shown in Figure \ref{fig:confmat-ahitype}, attaining a balanced accuracy of 80.11\%, weighted precision of 84.23\%, weighted recall of 80.70\%, weighted F1 of 80.88\%, MCC of 71.02\% and $\kappa$ of 70.16\%; when the problem is further simplified into 2-class detection where sleep apnea is diagnosed based on an AHI threshold, performance metrics significantly improve for various thresholds as demonstrated in Table \ref{tab:AHI_Table}.


\begin{table}[H]
    \centering
    \caption{Recording-wise test-set metrics for 2-class apnea prediction based on AHI thresholds.}
    \vspace{-3mm}
    \begin{tabular}{c|ccccccc}
        \hline
        \textit{Threshold} & \textit{Accuracy} & \phantom{\Large{A}}\textit{F1}\phantom{\Large{A}} & \textit{Precision} &  \textit{Recall} &  \textit{Specificity} & \textit{MCC} & $\kappa$ \\
        \hline
        \phantom{\Large{A}}5\phantom{\Large{A}} & 91.23 & 90.91 & 86.21 & 96.15 & 87.10 & 82.94 & 82.48  \\
        15 & 94.74 & 88.89 & 92.31 & 85.71 & 97.67 & 85.55 & 85.45 \\
        30 & 94.71 & 72.73 & 57.14 & 100.0 & 94.34 & 73.42 & 70.05 \\
        \hline
    \end{tabular}
    \label{tab:AHI_Table}
\end{table}
\FloatBarrier

\subsection{Segment-Wise Performance}

Table \ref{tab:Segment_Performace_Table} shows segment-wise metrics aggregated from the PSG sleep windows across all test set recordings for 1-second, 30-second, and 60-second windows. In spite of the severe class imbalance, the model reaches a MCC and $\kappa$ over 60\% and a balanced accuracy exceeding 81\% for all window sizes. The high F1, precision, and recall scores establish that this performance is not driven by simply predicting the majority class. Comparing the 30/60-second time windows to the 1-second predictions shows that there is an increase in MCC/$\kappa$/balanced accuracy and a decrease in specificity, w. F1, w. precision, and w. recall. The reduction is likely a consequence of max pooling (where 0 = no apnea, 1 = apnea) which labels a group of windows as apneas even if only a minority of windows were predicted apneas, inflating the number of false positives.

\begin{table}[H]
\begin{samepage}
    \centering
    \caption{Segment-wise test set performance metrics of the model on 1/30/60-second segments of sleep.}
    \vspace{-2mm}
    \resizebox{1\textwidth}{!}{
    \begin{tabular}{c|cccccccc}
        \hline
        \textit{Window Size (s)} & \textit{Balanced Accuracy} & \phantom{\Large{A}}\textit{W. F1}\phantom{\Large{A}} & \textit{W. Precision} &  \textit{W. Recall} &  \textit{Specificity} & \textit{MCC} & $\kappa$ &\textit{TAO} \\
        \hline
        1 & 81.81 & 94.70 & 94.81 & 94.61 & 96.79 & 61.48 & 61.43 & 76.11 \\
        30 & 84.85 & 92.76 & 92.85 & 92.69 & 95.45 & 68.34 & 68.32 & 73.60 \\
        60 & 85.45 & 91.85 & 91.94 & 91.78 & 94.71 & 69.68 & 69.66 & 73.91 \\
        \hline
    \end{tabular}
    }
    \label{tab:Segment_Performace_Table}
\end{samepage}
\end{table}

\subsection{Event-Wise Performance}

To evaluate event-wise metrics, we must define an IoU threshold: the minimum IoU two events must share in order to be considered a successful prediction. Setting the IoU threshold = 0.20 and computing all model performance metrics yields Table \ref{tab:EventWiseEval}.\\

\begin{table}[H]
    \caption{Event-wise evaluation of the proposed model under punitive and non-punitive conditions.}
    \vspace{-2mm}
    \resizebox{1\textwidth}{!}{
    \begin{tabular}{c|ccccc}
        \hline
        \textit{Method} & \textit{F1} & \textit{Recall/Sens.} & \textit{Precision} & \textit{OSI (fragments / event)} & \textit{USI (merges / event)} \\
        \hline
        \phantom{\Large{A}}Punitive\phantom{\Large{A}} & \makecell{69.33} & \makecell{71.98} & \makecell{66.87} & \makecell{0.0089} & \makecell{0.0269} \\
        Non-punitive & \makecell{71.30} & \makecell{74.66} & \makecell{68.23} & \makecell{0.0089} & \makecell{0.0269} \\
        \hline
    \end{tabular}
    }
    \label{tab:EventWiseEval}
\end{table}

This novel framework (detailed in section \ref{sec:statistical_analysis} and elaborated on in Appendix \ref{sec:appendix:eventwise_metrics}) enables the computation of numerous useful event-wise metrics, many of which are already common for segment-wise detection, such as recall, precision, and F1 score. Furthermore, these metrics are more intuitive and clinically applicable, seeing as they map directly onto detected vs. missed apnea episodes. For example, we can easily say that according to Table \ref{tab:EventWiseEval}, 72-75 percent of true apnea events are detected with a precision of 66-68 percent, depending on our fragmentation policy. Overall, an event-wise approach addresses limitations of traditional segment-wise classification, which relies on arbitrary time-segments and can overly penalize slight misalignments. \\

\subsection{Apnea Duration Analysis}

Apnea duration was computed by taking the length of each apnea interval which are at least 10 seconds for each recording in the test set, during periods of sleep detected by the sleep-wake model developed in \cite{zhang:2025}. The model achieves a moderate correlation for average apnea duration (left, $R=0.53$, $p < 0.01$, Mean Absolute Error (MAE) $= 6.71$ seconds) when compared to the PSG labels, as illustrated in Figure \ref{fig:avg_apnea_duration}. Furthermore, Figure \ref{fig:total_apnea_duration} in the Appendix shows that the model successfully captures the total apnea duration (left, $R=0.90$, $p < 0.001$, Mean Absolute Error (MAE) $= 413.39$ seconds) relative to the ground truth.

\begin{figure}[H]
    \centering
    \includegraphics[scale=0.5]{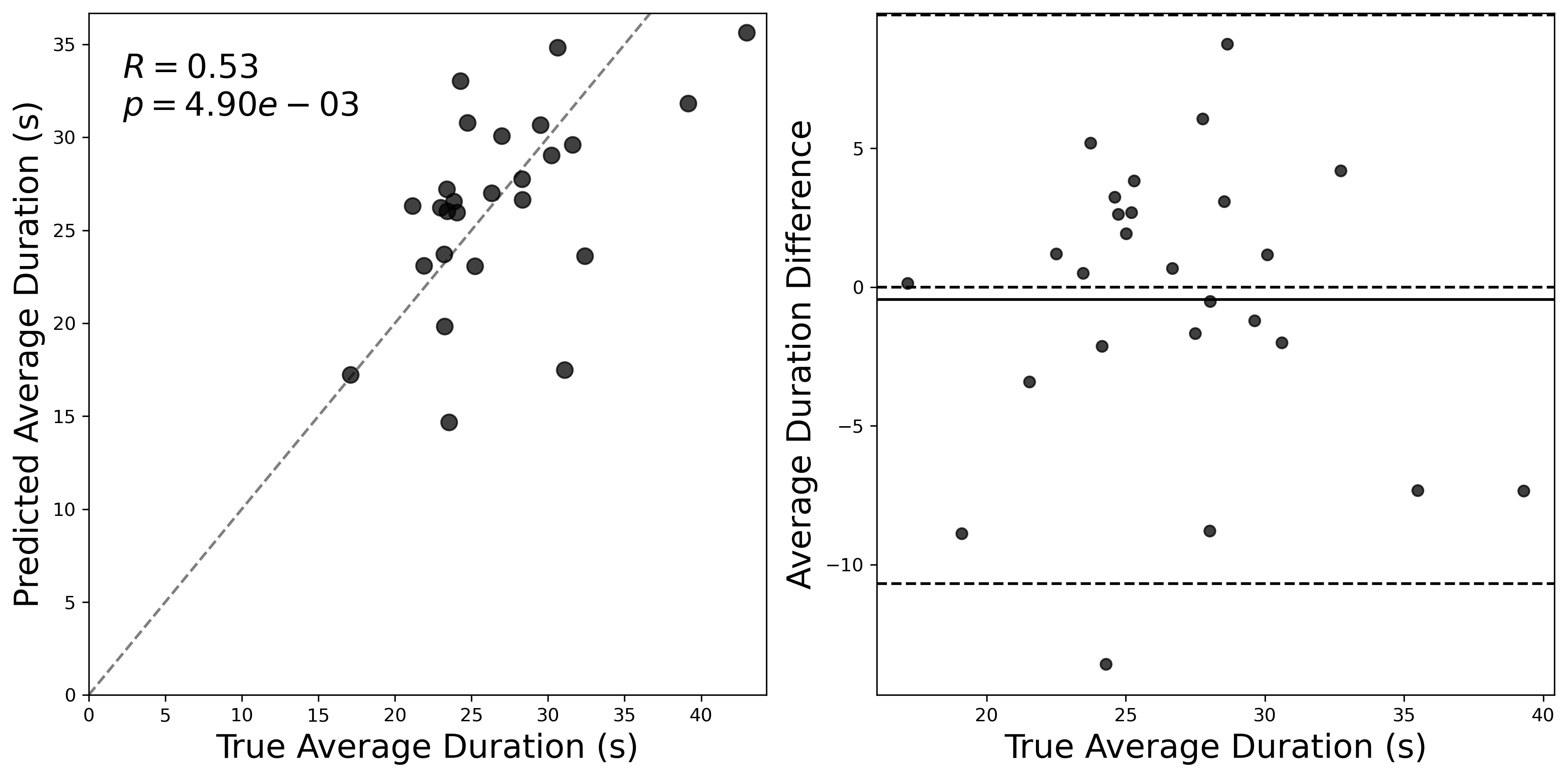}
    \vspace{-4mm}
    \caption{Scatterplot (left) and Bland-Altman plot (right) comparing the predicted and ground truth average apnea duration across the test set, when the ground truth AHI is greater than 5. For the Bland-Altman plot, the solid line is the bias and the dotted lines in the Bland-Altman plot represent 0 as well as 1.96 standard deviations above and below the mean.}
    \label{fig:avg_apnea_duration}
\end{figure}

\FloatBarrier

\subsection{Architecture Ablation Study}

An ablation study was conducted to determine the influence and importance of each component in the model architecture. Tables \ref{tab:ModelAblation} and \ref{tab:ModelAblation_4class} show segment-wise performance and recording-wise apnea severity predictions respectively for the proposed model compared to variants without a Context CNN and/or Distance MLP and with the bidirectional 4-layer Mamba block replaced by a bidirectional 4-layer LSTM. Postprocessing was not performed for models missing the distance MLP or the LSTM model.

\begin{table}[H]
    \caption{1-second window segment-wise evaluation metrics for architectures in the ablation study. Postprocessing is done on variants with the distance MLP. The best metric among variants is bolded.}
    \vspace{-2mm}
    \resizebox{1\textwidth}{!}{
    \begin{tabular}{c|ccccccc}
        \hline
        \textit{Architecture} & \textit{Balanced Accuracy} & \phantom{\Large{A}}\textit{W. F1}\phantom{\Large{A}} & \textit{W. Precision} &  \textit{W. Recall} &  \textit{Specificity} & \textit{MCC} & $\kappa$ \\
        \hline
        \phantom{\Large{A}}Original\phantom{\Large{A}} & \makecell{81.81} & \makecell{\textbf{94.70}} & \makecell{\textbf{94.81}} & \makecell{\textbf{94.64}} & \makecell{96.79} & \makecell{\textbf{61.48}} & \makecell{\textbf{61.43}} \\
        No Distance MLP & \makecell{\textbf{84.08}} & \makecell{93.62} & \makecell{94.50} & \makecell{93.07} &
        \makecell{94.61} &
        \makecell{58.11} & \makecell{57.04} \\
        No Context CNN & \makecell{81.05} & \makecell{94.42} & \makecell{94.46} & \makecell{94.11} &
        \makecell{96.33} &\makecell{58.88} & \makecell{58.71}\\
        No MLP \& No CNN & \makecell{78.30} & \makecell{94.14} & \makecell{94.14} & \makecell{94.14} & 
        \makecell{\textbf{96.84}} & \makecell{56.60} & \makecell{56.60} \\
        LSTM & \makecell{61.86} & \makecell{72.58} & \makecell{89.19} & \makecell{63.91} & 
        \makecell{64.26} &
        \makecell{12.72} & \makecell{8.14} \\
        \hline
    \end{tabular}
    }
    \label{tab:ModelAblation}
\end{table}

\begin{table}[H]
    \caption{4-class recording-level apnea severity evaluation metrics for architectures in the ablation study. Postprocessing is done on variants with the distance MLP. The best metric among variants is bolded.}
    \vspace{-2mm}
    \resizebox{1\textwidth}{!}{
    \begin{tabular}{c|ccccccc}
        \hline
        \textit{Architecture} & \textit{Balanced Accuracy} & \phantom{\Large{A}}\textit{W. F1}\phantom{\Large{A}} & \textit{W. Precision} &  \textit{W. Recall}  & \textit{MCC} & $\kappa$ \\
        \hline
        \phantom{\Large{A}}Original\phantom{\Large{A}} & \makecell{\textbf{80.11}} & \makecell{\textbf{80.88}} & \makecell{\textbf{84.23}} & \makecell{\textbf{80.70}} &  \makecell{\textbf{71.02}} & \makecell{\textbf{70.16}} \\
        No Distance MLP & \makecell{71.91} & \makecell{70.35} & \makecell{77.01} & \makecell{68.42} &
        \makecell{55.64} & \makecell{53.05} \\
        No Context CNN & \makecell{67.27} & \makecell{71.15} & \makecell{74.29} & \makecell{70.18} &\makecell{54.88} & \makecell{54.16}\\
        No MLP \& No CNN & \makecell{70.58} & \makecell{74.58} & \makecell{77.09} & \makecell{73.68} & \makecell{59.04} & \makecell{58.50} \\
        LSTM & \makecell{29.64} & \makecell{45.59} & \makecell{40.57} & \makecell{56.14} & 
        \makecell{15.95} & \makecell{11.76} \\
        \hline
    \end{tabular}
    }
    \label{tab:ModelAblation_4class}
\end{table}

As more components are removed from the model, all segment-wise metrics decline with the exception of balanced accuracy and specificity, while all 4-class metrics decline markedly, indicating these components are necessary for the model's performance. The increased balanced accuracy can be explained by the lack of fragmentation in predictions in favour of larger continuous apnea intervals leading to an over-prediction of apnea. The average apnea duration for the No Distance model is 29.17 seconds, while the Original model produces 27.23 seconds, causing the No Distance model to predict apnea for 11.15\% of the test set recordings, when the Original model and PSG predict 8.63\% and 7.82\% respectively. The model without the Distance MLP and Context CNN has the highest specificity, however the lower balanced accuracy signifies that the model struggles to find apnea intervals. The LSTM model significantly degrades in performance in both tasks but particularly for apnea severity. The degradation is due to the LSTM's tendency to predict extremely long apnea intervals, with an average apnea duration of 461.50 seconds (compared to 27.20 seconds for the ground truth), solidifying the advantage of using Mamba as the model backbone. Furthermore, postprocessing was not applied to the LSTM model because of its inability to predict distance values greater than 0. Given that a ReLU activation is applied at the end of the distance MLP to enforce distances to be greater than or equal to 0, the LSTM consistently converges to the trivial all-zero solution. This suggests that the LSTM backbone provides less meaningful representations compared to its Mamba counterpart under the same conditions.


\subsection{Feature Ablation Study}

To determine the significance of the sets of sensors provided by the ANNE One device, a feature ablation study was performed comparing the chest module (ECG and XYZ) to the limb module (PPG). Because ECG and PPG are most likely to experience poor signal quality, a model trained only on accelerometry data was also created. As illustrated by Table \ref{tab:FeatureAblation}, while the chest module is the most informative component, the limb module trails closely behind and is superior at filtering out false positives. The model trained without ECG or PPG information outperforms the chest-only and limb-only models in balanced accuracy but is worse for most other metrics, indicating that accelerometry alone insufficient to distinguish apnea events and noisy arousals. 

\begin{table}[H]
    \caption{1-second window segment-wise performance metrics for feature sets in the ablation study. The best metric among variants is bolded.}
    \vspace{-2mm}
    \resizebox{1\textwidth}{!}{
    \begin{tabular}{c|cccccccc}
        \hline
        \textit{Feature Set} & \textit{Balanced Accuracy} & \phantom{\Large{A}}\textit{W. F1}\phantom{\Large{A}} & \textit{W. Precision} &  \textit{W. Recall} &  \textit{Specificity} & \textit{MCC} & $\kappa$ & TAO  \\
        \hline
        
        \phantom{\Large{A}}Full\phantom{\Large{A}} & \makecell{\textbf{81.81}} & \makecell{\textbf{94.70}} & \makecell{\textbf{94.81}} & \makecell{\textbf{94.64}} & \makecell{96.79} & \makecell{\textbf{61.48}} & \makecell{\textbf{61.43}} & \makecell{\textbf{76.11}} \\
        Chest Module &  \makecell{78.05} & \makecell{93.39} & \makecell{93.65} & \makecell{93.17} & \makecell{95.76} & \makecell{52.77} & \makecell{52.63} & \makecell{69.87}\\
        Limb Module & \makecell{74.11} & \makecell{93.46} & \makecell{93.33} & \makecell{93.60} & \makecell{\textbf{96.93}} & \makecell{50.52} & \makecell{50.45} & \makecell{60.21}\\
        No ECG or PPG & \makecell{79.88} & \makecell{92.85} & \makecell{93.61} & \makecell{92.31} & \makecell{94.42} & \makecell{51.92} & \makecell{51.18} & \makecell{75.00}\\
        \hline
    \end{tabular}
    }
    \label{tab:FeatureAblation}
\end{table}

\subsection{Performance By Clinical Group}
After MAE for each AHI prediction was calculated, linear regression was performed of AHI MAE against clinical variables to determine the model's robustness. There is no significant corelation between sex and PLMS with AHI MAE. Although the model obtains higher MAE for older patients ($p = 0.046$), the trend loses significance ($p = 0.098$, $R = 0.25$) when five non-apnea patients (those with true and predicted AHI values of 0) are excluded, suggesting that the correlation with age is mostly driven by a shortage of apnea in younger testing set patients.\\ 



There were no differences in AHI MAE (t-test, $t=-0.23$, $p=0.81$) based on the identity of the 2 technicians who manually labelled the recordings in the testing set.

\begin{figure}[H]
    \centering
    \includegraphics[scale=0.45]{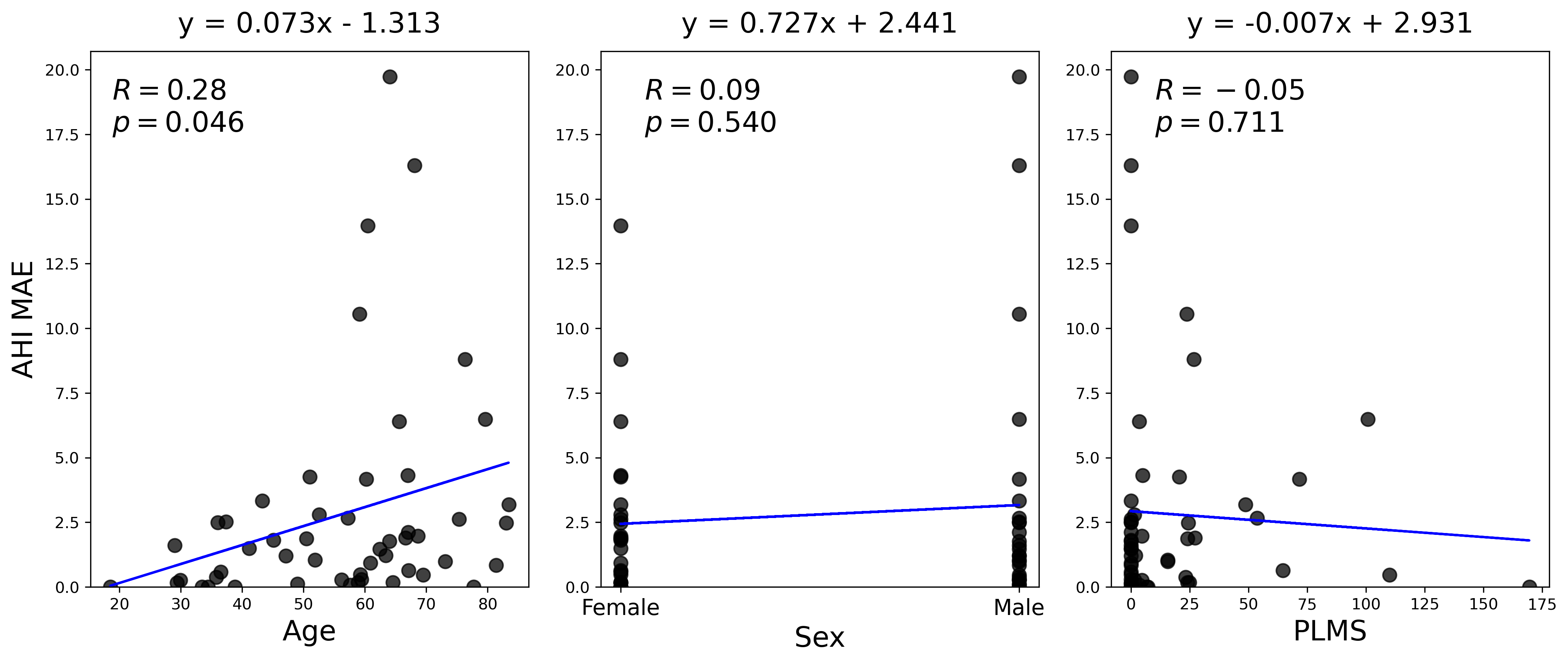}
    \vspace{-2mm}
    \caption{Scatter plots of MAE for recording wise AHI prediction against different demographic groups. A linear regression line is plotted, with its equation above.}
    \label{fig:demographic_scatterplots}
\end{figure}




\subsection{Signal Quality Analysis}

The model's robustness to noise was examined by measuring segment-wise performance for 1-second time windows where some combination of ECG and PPG signals experienced failure. As shown in table \ref{tab:SQI_Analysis_Table}, performance metrics decline as more signals experience poor quality but remain considerably high even even when both signals fail, implying that other signals like accelerometry, and the features derived from them, can capture some of the information lost in the ECG and PPG. The bad-PPG metrics lag notably behind the bad-ECG metrics, indicating that PPG is the more consequential of the two signals. 


\begin{table}[H]
\begin{samepage}
    \centering
    \caption{1-second window segment-wise performance metrics on the test set, stratified by which combination of ECG/PPG signals experienced poor quality. The number and percentage of time windows containing an apnea are tabulated on the right.}
    \vspace{-2mm}
    \resizebox{1\textwidth}{!}{
    \begin{tabular}{c|ccccccc|cc}
        \hline
        \textit{Signal Failure} & \textit{Balanced Accuracy} & \phantom{\Large{A}}\textit{W. F1}\phantom{\Large{A}} & \textit{W. Precision} &  \textit{W. Recall} &  \textit{Specificity} & \textit{MCC} & $\kappa$ & \textit{\# Apnea} & \textit{\% Apnea} \\
        \hline
        None & 82.97 & 94.79 & 94.97 & 94.64 & 96.62 & 62.47 & 62.34 & 60696 & 7.21\\
        ECG & 78.97 & 93.98 & 93.92 & 94.05 & 96.99 & 59.35 & 59.32 & 10505 & 8.15\\
        PPG & 68.21 & 94.17 & 93.93 & 94.47 & 97.70 & 40.83 & 40.51 & 2734 & 5.49\\
        ECG or PPG & 76.93 & 94.08 & 93.98 & 94.20 & 97.21 & 56.11 & 56.05 & 12766 & 7.42 \\
        ECG and PPG & 72.63 & 93.40 & 93.28 & 93.53 & 96.84 & 47.15 & 47.10 & 473 & 6.83\\
        \hline
    \end{tabular}
    }
    \label{tab:SQI_Analysis_Table}
\end{samepage}
\end{table}

\begin{figure}[H]
    \centering
    \includegraphics[scale=0.35]{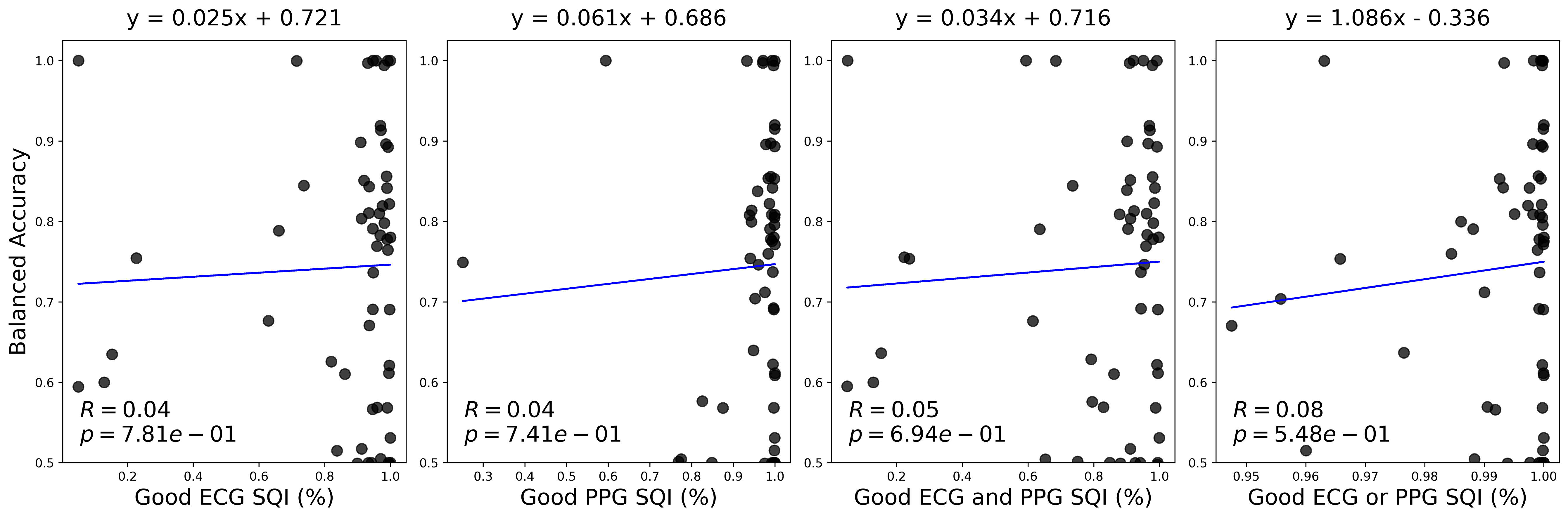}
    \vspace{-2mm}
    \caption{Scatter plots of the recording wise balanced accuracy against the proportion of high ECG SQI, PPG SQI, both or either in each recording. A linear regression line is plotted, with its equation above.}
    \label{fig:sqi_scatterplots}
\end{figure}

\subsection{Multiclass Classification}

\begin{figure}[H]
    \centering
    \centerline{\includegraphics[width=1.1\textwidth]{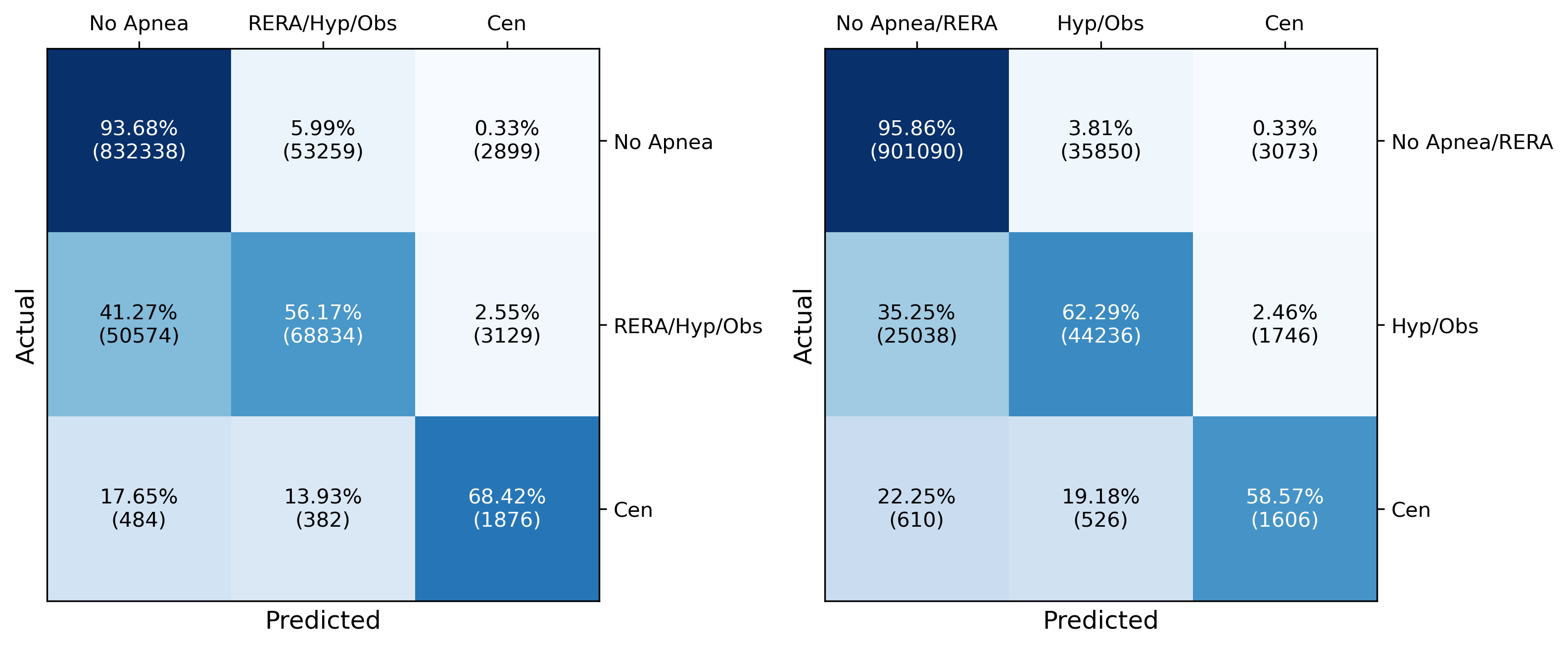}}
    \vspace{-2mm}
    \caption{Confusion matrices for segment-wise multiclass apnea classification of models trained to classify non-apneas, RERAs/hypopneas/obstructive apneas, and central apneas (left); as well as non-apneas/RERAs, hypopneas/obstructive apneas, and central apneas (right).}
    \label{fig:multiclass-apnea}
\end{figure}

Multiclass classification was performed by training models on classes No Apnea, Hypopnea/Obstructive Apnea, and Central Apnea, with RERA events being combined with both the No Apnea class and the Hypopnea/Obstructive Apnea class. Figure \ref{fig:multiclass-apnea} presents a three class confusion matrix, with classes No Apnea, Obstructive Apnea, which contains RERA, Hypopnea and Obstructive Apnea events, and Central Apnea. In addition, Table \ref{tab:multiclass-apnea} provides segment wise performance over 1, 30 and 60 segments of sleep, where the 30 and 60 second windows were determined by a majority vote. The confusion between the No Apnea and RERA/Hyp./Obs. class stems from the inability to distinguish RERAs from normal events, which is typically performed on a signal which the ANNE One device does not have. \\

\begin{figure}[H]
    \centering
    \centerline{\includegraphics[width=0.9\textwidth]{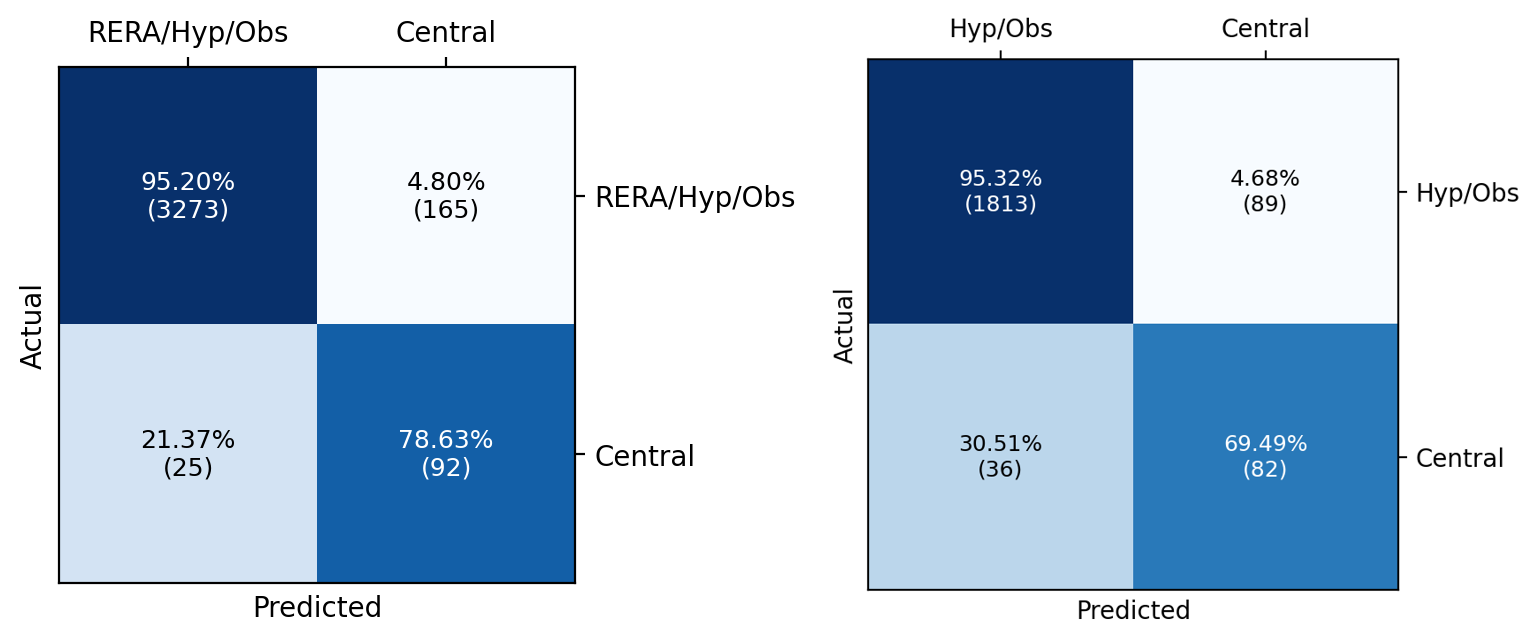}}
    \vspace{-2mm}
    \caption{Confusion matrices for event-type multiclass apnea classification.}
    \label{fig:eventwise-multiclass-confmat}
\end{figure}

An event-wise evaluation of the multiclass model(s) was conducted to assess classification performance across event types (event-type analysis). Figure \ref{fig:eventwise-multiclass-confmat} and Table \ref{tab:multiclass-eventwise} present the results of event-type classification, which measures how accurately the model labeled events that were successfully detected. For example, among the 3,555 detected events, 95\% of RERA/Hyp/Obs and 77\% of Central Apnea events were correctly classified. \\

\begin{table}[H]
\begin{samepage}
    \centering
    \caption{X-second window segment-wise test set performance metrics for two multi-class apnea detection models, on 1/30/60-second segments of sleep.}
    \vspace{-2mm}
    \resizebox{1\textwidth}{!}{
    \begin{tabular}{c|c|ccccccc}
    \hline
        \textit{Model} & \textit{Window Size (s)} & \textit{Balanced Accuracy} & \phantom{\Large{A}}\textit{W. F1}\phantom{\Large{A}} & \textit{W. Precision} &  \textit{W. Recall} &  \textit{Specificity} & \textit{MCC} & $\kappa$ \\
        \hline
        \multirow{3}{*}{RERA/Hyp/Obs} & 1  & 72.76  & 89.22  & 89.43 & 89.08 & 93.68 & 50.74 & 50.72  \\
                             & 30 & 73.80 & 89.24 & 89.21 & 89.40 & 94.70 & 48.87 & 48.80  \\
                             & 60 & 74.87 & 90.38  & 90.34 & 90.55 & 95.41 & 49.00 & 48.90  \\
        \hline
        \multirow{3}{*}{Hyp/Obs} & 1  & 72.24  & 93.70  & 94.07 & 93.41 & 95.85 & 55.29 & 55.08  \\
         & 30 & 73.66 & 93.61 & 93.81 & 93.45 & 96.23 & 54.30 & 54.24  \\
         & 60 & 73.99  & 93.64 & 93.83 & 93.51 & 96.28 & 52.41 & 52.37  \\
        \hline
    \end{tabular}
    }
    \label{tab:multiclass-apnea}
\end{samepage}
\end{table}

\begin{table}[H]
\begin{samepage}
    \centering
    \caption{Event-type performance metrics summarized.}
    \vspace{-2mm}
    \resizebox{1\textwidth}{!}{
    \begin{tabular}{c|ccccccc}
        \hline
        \textit{Model} & \textit{Accuracy} & \textit{Balanced Accuracy} & \textit{W. Precision} & \textit{W. Recall / Sensitivity} & \textit{W. F1} & \textit{MCC} & $\kappa$ \\
        \hline
        RERA/Hyp/Obs & 94.7 & 86.9 & 97.2 & 94.7 & 95.6 & 50.9 & 46.8 \\
        Hyp/Obs      & 93.8 & 82.4 & 95.1 & 93.8 & 94.3 & 54.6 & 53.5 \\
        \hline
    \end{tabular}
    }
    \label{tab:multiclass-eventwise}
\end{samepage}
\end{table}

\section{Discussion}

\subsection{Key Findings}
This study has established the ability of a Mamba-based model with auxiliary labels, such as the distance metric, to accurately classify apnea intervals using the ANNE One wearable device without oxygen flow information. In addition, the model can accurately classify the recordings according to specific AHI bins and thresholds. The model shows good performance on a diverse clinical dataset, generalizing to a population with both apnea and non-apnea related issues.

\subsection{Comparison to Human Classification}
In a recent study conducted by C Davies, JY Lee, J Walter, et al., researchers evaluated the ANNE One by determining how well a human specialist can classify recordings generated by the wearable device compared to its corresponding PSG classification \cite{anne_baseline:2022}. The study calculated similar recording-wise metrics, including a scatterplot and statistics for AHI cutoffs at 5, 15 and 30, similar to Table \ref{tab:AHI_Table} \cite{anne_baseline:2022}. When determining the correlation between the predicted AHI computed on the ANNE One device compared to the PSG baseline, the researchers obtained a correlation of $R=$0.93 \cite{anne_baseline:2022}. Our model is able to surpass this correlation coefficient with a value of $R=$ 0.95 (Figure \ref{fig:scatter-blandaltman}), showing a stronger level of agreement with the ground-truth PSG. \\

Furthermore, the study generated results classifying recordings based on AHI thresholds 5, 15 and 30 \cite{anne_baseline:2022}. As seen in Table \ref{tab:human_model_rec_comp}, the current model is able to match or, in some cases exceed the performance of human specialists. Having the ability to correctly distinguish recordings using key thresholds for quick and efficient OSA detection is crucial, and the model can generate similar results to a human specialist in a fraction of the time.

\begin{table}[H]
    \centering
    \caption{Comparison between current model and human specialists for classifying AHI on the ANNE One device \cite{anne_baseline:2022}.}
    \begin{tabular}{c|cccc}
    \hline
    \textit{Model} & \textit{Threshold} & \textit{Accuracy} & \textit{Sensitivity} & \textit{Specificity} \\
    \hline
    \multirow{3}{*}{Current Model} & 5  & 91.23  & 96.15  & 87.10  \\
                         & 15 &  94.74 & 85.71 & 97.67  \\
                         & 30 & 94.71  & 100.0 & 94.34  \\
    \hline
    \multirow{3}{*}{Human Specialists} & 5  & 92.00  & 96.00  & 86.00 \\
                         & 15 & 95.00  & 90.00  & 97.00  \\
                         & 30 & 94.00  & 65.00  & 99.00  \\
    \hline
    \end{tabular}
    \label{tab:human_model_rec_comp}
\end{table}
\FloatBarrier

\subsection{Comparison With Related Studies}























OSA detection methods using machine learning have been trained on datasets from wearable devices (Table \ref{tab:related-wearable}) and PSGs (Table \ref{tab:related-psg}), with models in latter group typically achieving superior performance than the former in segment-wise tasks due to good signal quality. Even still, our model exceeds those trained from wearable devices who provided segment-wise results (\cite{Chen:2024}, \cite{ZouAndLiu:2024}, \cite{Yeo:2022}), in addition to outperforming \cite{Retamales:2024}, \cite{Kwon:2023}, \cite{Sharma:2022}, and \cite{Yeo:2022} for PSG trained models. The ones that our model trail (\cite{mamba_apnea:2025}, \cite{Zhou:2024}) are trained on PSG data, specifically the Apnea-ECG dataset, which would be considered cleaner and less noisy. As a result, the Mamba-based model (\cite{mamba_apnea:2025}) does better than our model in terms of segment-wise performance, however, we are able to match the recording-wise performance (apart from the AHI being less than 5), despite the differences in dataset quality. Compared to models trained on wearable device data, we outperform or match performance for all AHI thresholds in recording-wise tasks, while having the highest segment-wise metrics. Unlike most other papers in Table \ref{tab:related-wearable}, we have consistently high recording-wise performance across different AHI thresholds, and comparable metrics to PSG trained models in Table \ref{tab:related-psg} despite having a larger, noisier dataset in most cases. In addition to some related literature not including segment-wise performance, or various cutoffs for recording-wise performance, none consider other important analyses such as event-wise, event-type and average apnea duration performance. \\


\begin{table}[H]
    \caption{Automatic OSA detection models in related literature that are trained (unless otherwise specified) and evaluated on wearable data. $n$ is the number of recordings in the training, validation, and the row-specific test sets. SW = Segment-wise performance (60-second windows with unweighted metrics), RW($x$) = Recording-wise performance with AHI threshold $x$. EW = Event-Wise performance (punitive). AAD = Average Apnea Duration Performance. ET = Event Type Performance.}
    \vspace{-2mm}
    \resizebox{1\textwidth}{!}{
    \begin{tabular}{c|ccccccccc}
        \hline
        \textit{Study} & \textit{Signals} & \textit{Dataset} & \textit{SW} & \textit{RW (5)} & \textit{RW (15)} & \textit{RW (30)} & \textit{EW} & \textit{AAD} & \textit{ET} \\
        \hline
        
        \makecell{This Study\\(2025)} & \makecell{ECG,\\ PPG,\\ Accelerometry} & \makecell{Sunnybrook Health\\Sciences Center\\ ($n=384$)} & \makecell{Acc: 91.78 \\ Sens: 76.18 \\ Spec: 94.71}  & \makecell{Acc: 91.23 \\ Sens: 96.15 \\ Spec: 87.10} & \makecell{Acc: 94.74 \\ Sens: 85.71 \\ Spec: 97.67} & \makecell{Acc: 94.71 \\ Sens: 100.0 \\ Spec: 94.34} & \makecell{F1: 72.13 \\ Sens: 75.59 \\ Prec: 68.98} & \makecell{R: 0.53 \\ $p < 0.05$ \\MAE: 6.71} & \makecell{W. F1: 0.956 \\ W. Sens: 0.947 \\ W. Prec: 0.972}\\
        \hline

        \makecell{Dang et al.\\(2025) \cite{Dang:2025}} & \makecell{Pulse Oximetry,\\Thoracic/Abdominal\\Movement,\\Accelerometry} & \makecell{Ulsan National Institute\\of Science and\\Technology\\($n=35$)} & \makecell{N/A}  & \makecell{Acc: 85.81\\Sens: 90.91\\Spec: 76.92} & \makecell{Acc: 97.14\\Sens: 100.0\\Spec: 95.45} & \makecell{Acc: 97.14\\Sens: 100.0\\Spec: 96.00} & N/A & N/A & N/A \\
        \hline

        \makecell{Goldstein et al.\\(2025) \cite{Goldstein:2025}} & \makecell{"10 physiological \\channels including ECG"} & \makecell{7 Locations\\($n=340$)} & \makecell{N/A}  & \makecell{Acc: 92.09\\Sens: 94.12\\Spec: 88.39} & \makecell{Acc: 91.46\\Sens: 89.17\\Spec: 92.86} & \makecell{Acc: 96.52\\Sens: 91.04\\Spec: 97.99} & N/A & N/A & N/A \\
        \hline

        \makecell{Hayano et al.\\(2025) \cite{Hayano:2025}} & \makecell{Inertial Measurement\\Unit, Gyroscope,\\Accelerometry} & \makecell{Akita University Hospital,\\Medical Corporation\\Sound Sleep,\\Medical Corporation \\Zuimeikai\\($n=126$)} & \makecell{N/A}  & \makecell{Acc: 84.09\\Sens: 97.30\\Spec: 14.29} & \makecell{Acc: 86.36\\Sens: 92.31\\Spec: 77.78} & \makecell{Acc: 95.45\\Sens: 100.0\\Spec: 93.33} & N/A & N/A & N/A \\
        \hline

        \makecell{Chen et al.\\(2024) \cite{Chen:2024}} & \makecell{PPG} & \makecell{Affiliated Hospital of\\Sun Yat-sen University\\ ($n=92$)} & \makecell{Acc: 74.08 \\ Sens: 50.99 \\ Spec: 84.04}  & \makecell{N/A} & \makecell{N/A} & \makecell{Acc: 85.79 \\ Sens: 64.00 \\ Spec: 92.53} & N/A & N/A & N/A \\
        \hline

        \makecell{Zou \& Liu \\(2024) \cite{ZouAndLiu:2024}} & \makecell{PPG} & \makecell{Affiliated Hospital of\\Sun Yat-sen University\\ ($n=92$)} & \makecell{Acc: 82.76 \\ Sens: 71.58 \\ Spec: 86.74}  & \makecell{N/A} & \makecell{N/A} & \makecell{Acc: 97.83 \\ Sens: 88.89 \\ Spec: 100.0} & N/A & N/A & N/A \\
        \hline
        
        \makecell{Strumpf et al.\\(2023) \cite{Strumpf:2023}} & \makecell{SPO2,\\ Pulse Oximetry,\\
        Accelerometry} & \makecell{Cleveland Medical Center\\Bolwell and Beachwood \\Sleep Labs\\($n=84$)} & \makecell{N/A}  & \makecell{Acc: 85 \\ Sens: 92 \\ Spec: 64} & \makecell{Acc: 89 \\ Sens: 91 \\ Spec: 88} & \makecell{Acc: 91 \\ Sens: 83 \\ Spec: 93} & N/A & N/A & N/A \\
        \hline

        \makecell{Xu et al.\\(2023) \cite{Xu:2023}} & \makecell{SPO2,\\Heart Rate,\\Body Movement} & \makecell{Guangdong Provincial\\People's Hospital\\($n=196$)} & \makecell{N/A}  & \makecell{Acc: 91\\Sens: 93\\Spec: 89} & \makecell{Acc: 91\\Sens: 92\\Spec: 89} & \makecell{N/A} & N/A & N/A & N/A \\
        \hline

        \makecell{Zhou et al.\\(2023) \cite{Zhou:2023}} & \makecell{PPG, Audio,\\Questionnaire,\\Accelerometry} & \makecell{Shenzhen People's Hospital\\($n=350$)} & \makecell{N/A}  & \makecell{Acc: 88.1\\Sens: 89.1\\Spec: 75.5} & \makecell{Acc: 84.5\\Sens: 84.2\\Spec: 85.3} & \makecell{Acc: 85.1\\Sens: 85.6\\Spec: 84.7} & N/A & N/A & N/A \\
        \hline

        \makecell{Yeo et al.\\(2022) \cite{Yeo:2022} \\ \textit{From Table \ref{tab:related-psg}}} & \makecell{ECG, \\ Accelerometry} & \makecell{Kyunghee University\\Hospital\\($n=340$)} & \makecell{Acc: 84 \\ Sens: 71 \\ Spec: 88}  & \makecell{N/A} & \makecell{Acc: 92 \\ Sens: 88 \\ Spec: 95} & \makecell{N/A} & N/A & N/A & N/A \\
        \hline

        \makecell{Yeh et al.\\(2021) \cite{Yeh:2021}} & \makecell{PPG,\\Pulse Oximetry,\\Accelerometry} & \makecell{University Hospitals\\Cleveland Medical Center\\($n=336$)} & \makecell{N/A}  & \makecell{Acc: 56.4\\Sens: 100.0\\Spec: 2.9} & \makecell{Acc: 80.8\\Sens: 93.1\\Spec: 73.5} & \makecell{Acc: 91.0\\Sens: 71.4\\Spec: 95.3} & N/A & N/A & N/A \\
        \hline


        \makecell{Hafezi et al.\\(2020) \cite{Hafezi:2020}} & \makecell{Accelerometry} & \makecell{Toronto Rehabilitation\\Institute\\($n=69$)} & \makecell{N/A}  & \makecell{Acc: 78\\Sens: 98\\Spec: 36} & \makecell{Acc: 84\\Sens: 81\\Spec: 87} & \makecell{Acc: 88\\Sens: 67\\Spec: 94} & N/A & N/A & N/A \\
        \hline

        \makecell{Pillar et al.\\(2020) \cite{Pillar:2020}} & \makecell{Pulse Arrival Time} & \makecell{11 Locations\\($n=84$)} & \makecell{N/A}  & \makecell{N/A} & \makecell{Acc: 88.3\\Sens: 71.4\\Spec: 98.6} & \makecell{N/A} & N/A & N/A & N/A \\
        \hline
    \end{tabular}
    }
    \label{tab:related-wearable}
\end{table}





\begin{table}[H]
    \caption{Automatic OSA detection models in related literature that are trained (unless otherwise specified) and evaluated on PSG data. $n$ is the number of recordings in the training, validation, and the row-specific test sets. SW = Segment-wise performance (60-second windows with unweighted metrics), RW ($x$) = Recording-wise performance with AHI threshold $x$.}
    \vspace{-2mm}
    \resizebox{1\textwidth}{!}{
    \begin{tabular}{c|cccccc}
        \hline
        \textit{Study} & \textit{Signals} & \textit{Dataset} & \textit{SW} & \textit{RW (5)} & \textit{RW (15)} & \textit{RW (30)}\\
        \hline

        Li et al. (2025) \cite{mamba_apnea:2025} & \makecell{ECG} & \makecell{Apnea-ECG\\($n=70$)} & \makecell{Acc: 91.91 \\ Sens: 89.00 \\ Spec: 93.70}  & \makecell{Acc: 100.0 \\ Sens: 100.0 \\ Spec: 100.0} & \makecell{Acc: 91.43\\Sens: 100.0\\Spec: 82.35} & \makecell{Acc: 94.29\\Sens: 85.71\\Spec: 100.0} \\
        \hline

        Li et al. (2025) \cite{mamba_apnea:2025} & \makecell{ECG} & \makecell{The Second Affiliated Hospital\\of Xi’an Jiaotong University\\($n=185$)} & \makecell{Acc: 82.95\\Sens: 76.60\\Spec: 90.48} & \makecell{N/A} & \makecell{N/A} & \makecell{N/A} \\
        \hline

        \makecell{Retamales et al. (2024) \cite{Retamales:2024}} & \makecell{Pulse Oximetry,\\ Thoracic/Abdominal\\Movement} & \makecell{MESA, Sleep Heart Health Study,\\Osteroporotic Fractures in Men\\ ($n=10643$)} & \makecell{Acc: 86*\\Sens: 67*\\Spec: 91*}  & \makecell{Acc: 92.23\\Sens: 96.62\\Spec: 59.11} & \makecell{Acc: 88.01\\Sens: 86.35\\Sens: 90.10} & \makecell{Acc: 91.35\\Sens: 74.11\\Spec: 97.36} \\
        \hline

        \makecell{Zhou \& Kang (2024) \cite{Zhou:2024}} & \makecell{ECG} & \makecell{Apnea-ECG\\ ($n=70$)} & \makecell{Acc: 96.37\\Sens: 95.67\\Spec: 97.44}  & \makecell{N/A} & \makecell{N/A} & \makecell{N/A} \\
        \hline

        \makecell{Kwon et al. (2023) \cite{Kwon:2023}} & \makecell{EOG, EEG, EMG} & \makecell{University of Coimbra,\\Unspecified\\ ($n=72$)} & \makecell{Acc: 88.52\\Sens: 24.76\\Spec: 99.69}  & \makecell{N/A} & \makecell{N/A} & \makecell{N/A} \\
        \hline

        \makecell{Sharma et al. (2022) \cite{Sharma:2022}} & \makecell{Pulse Oximetry} & \makecell{University College Dublin\\ ($n=25$)} & \makecell{Acc: 89.21\\Sens: 92.34\\Spec: 89.13}  & \makecell{N/A} & \makecell{N/A} & \makecell{N/A} \\
        \hline

        Yeo et al. (2022) \cite{Yeo:2022} & \makecell{ECG, \\ Accelerometry} & \makecell{Cleveland Family Study\\($n=326$)} & \makecell{Acc: 85 \\ Sens: 68 \\ Spec: 89}  & \makecell{N/A} & \makecell{Acc: 88 \\ Sens: 81 \\ Spec: 90} & \makecell{N/A} \\
        \hline

        Yeo et al. (2022) \cite{Yeo:2022} & \makecell{ECG, \\ Accelerometry} & \makecell{Multi-Ethnic Study\\of Atherosclerosis (MESA)\\($n=1104$)} & \makecell{Acc: 80 \\ Sens: 66 \\ Spec: 84}  & \makecell{N/A} & \makecell{Acc: 81 \\ Sens: 80 \\ Spec: 82} & \makecell{N/A} \\
        \hline


    \end{tabular}
    }
    \label{tab:related-psg}
    \caption*{\footnotesize *The metric was calculated on 30-second windows and approximated based on Figure 2 inside the paper.}
\end{table}
\FloatBarrier

\subsection{Architecture/Model Analysis}

To further examine the quality of model performance, the average and total apnea duration for each recording was computed. Apnea duration is an important metric in clinical studies, therefore a strong correlation is crucial as it distinguishes the model's ability to capture the relevance of each apnea \cite{ButlerMatthewP.2019AEDP}. The moderate correlation for average apnea duration (R=0.53) coupled with the strong correlation for the total duration (R=0.90), demonstrates the model's ability to effectively capture event duration.\\

The model is shown to be generalizable across clinical subgroups (Sex, PLMS) with $p > 0.05$ when evaluating on MAE, except for Age. The moderate correlation and significant $p$-value for age can be explained by the lack of apnea in younger subjects. \\ 

The feature ablation study establishes that the model relies heavily on the limb module (PPG, SPO2) to differentiate apnea intervals from non-related arousals, established by the high specificity shown in Figure \ref{tab:FeatureAblation}. The high specificity is likely due to the inclusion of the SPO2 feature, which is crucial for OSA detection without oxygen flow information. However, the limb module alone is not able to capture all of the apnea intervals, exemplified by the low balanced accuracy (74.11\%) and a TAO of 60.21\%. The 4\% increase in balanced accuracy between the limb and chest module (ECG and XYZ), and the substantial increase in TAO (69.87\%) demonstrates the necessity of ECG and accelerometry data to detect apnea related arousals. Furthermore, using only accelerometry data (XYZ, body roll, body pitch) without PPG and ECG information is able to achieve a higher TAO (75.00\%) compared to both the chest and limb modules. This highlights the richness of the features provided by the accelerometer, in addition to the benefits of having the device located on the chest. \\


To extend upon current research and to test the validity of the current architecture, the model was trained in a multiclass setting. Specifically, the model was trained on classes No Apnea, RERA/Hypopnea/ Obstructive Apnea and Central Apnea, in addition to classes No Apnea/RERA, Hypopnea/Obstructive Apnea and Central Apnea. Of note, the most difficult class to differentiate is the RERA class, due to the airflow instruments being unavailable to the ANNE One device.

\subsection{Improving True Apnea Overlap}
Although the model achieves high accuracy and high specificity for AHI classification, it may also be helpful to focus on maximizing TAO, the proportion of apneas overlapping with a prediction. Although the model might not yet be able to fully annotate a recording, a practical clinical workflow could include a model that generally identifies 90+\% of apneas, followed by a technician going through the model's predictions and adjusting boundaries or removing false positives. This saves the technician the time it would take to screen an entire recording, allowing them to only focus on the events detected by the model. \\

From this perspective, practical implementations could focus on maximizing TAO, even at the cost of specificity. One way to improve TAO is to adjust the post-processing described in Section \ref{sec:postprocessing}. Currently, any predictions shorter than 10 seconds are removed, eliminating lower-confidence predictions that last only a few seconds. By removing this postprocessing step, more predictions can be made, and TAO increases from 77.11 to 84.66, while specificity decreases from 96.79 to 96.19. \\

\subsection{Limitations and Future Work}

The proposed model has many strengths: accurate classification of recordings based on AHI thresholds, high accuracy for segment-wise apnea classification, and robustness to changes in demographics. Despite these strengths, the model still struggles to capture the exact boundaries of an apnea interval. As indicated by the event-wise metrics and TAO, the model predicts apnea in generally the correct locations (capturing over 75\% of events). However, as evidenced by the analysis in Appendix \ref{sec:appendix:eventwise_metrics} and by the duration analysis, the model tends to over or under-predict the durations of apnea intervals, or makes predictions offset by a few seconds. Therefore, additional work is necessary to develop a model to accurately predict these boundaries. Furthermore, given the recordings were all obtained on a single site, additional work is needed to determine the generalizability of the model when tested on recordings from a different setting.

\section{Conclusion}
This study introduces a Mamba-based neural network with an auxiliary distance metric to perform binary and multiclass classification automatically using data generated by the ANNE One wearable device, which does not possess an oxygen flow signal. The model is able to surpass or meet metrics in related literature when applied to our clinically diverse, ANNE One dataset. In addition to addressing multiclass classification for apnea, which is underrepresented in current literature, this research introduces new metrics for a comprehensive analysis of apnea detection models such as an apnea duration analysis, event-wise/event-type performance and TAO. 

\section*{Acknowledgments}
This study was not funded by the manufacturer of the ANNE One Device, Sibel Health.

\section*{Data Availability}
Upon publication, all data will be made available at \hyperlink{https://www.frdr-dfdr.ca/repo/}{https://www.frdr-dfdr.ca/repo/}. Furthermore, all code will be uploaded to \hyperlink{https://github.com/OSHS2019/}{https://github.com/OSHS2019/}.

\newpage
\printbibliography 

\newpage

\appendix
\section*{Appendix}
\section{Model Parameters} \label{model-params}

Below is the printed output of the binary apnea classification model with exact hyperparameter values.
\begin{lstlisting}[basicstyle=\footnotesize]
MambaNet(
  (mamba_forward): Mamba(
    (layers): ModuleList(
      (0-3): 4 x ResidualBlock(
        (mixer): MambaBlock(
          (in_proj): Linear(in_features=75, out_features=300, bias=True)
          (conv1d): Conv1d(150, 150, kernel_size=(4,), stride=(1,), padding=(3,), groups=150)
          (x_proj): Linear(in_features=150, out_features=37, bias=False)
          (dt_proj): Linear(in_features=5, out_features=150, bias=True)
          (out_proj): Linear(in_features=150, out_features=75, bias=True)
        )
        (norm): RMSNorm()
      )
    )
  )
  (mamba_backward): Mamba(
    (layers): ModuleList(
      (0-3): 4 x ResidualBlock(
        (mixer): MambaBlock(
          (in_proj): Linear(in_features=75, out_features=300, bias=True)
          (conv1d): Conv1d(150, 150, kernel_size=(4,), stride=(1,), padding=(3,), groups=150)
          (x_proj): Linear(in_features=150, out_features=37, bias=False)
          (dt_proj): Linear(in_features=5, out_features=150, bias=True)
          (out_proj): Linear(in_features=150, out_features=75, bias=True)
        )
        (norm): RMSNorm()
      )
    )
  )
  (norm1): BatchNorm1d(150, eps=1e-05, momentum=0.1, affine=True, track_running_stats=True)
  (norm2): BatchNorm1d(96, eps=1e-05, momentum=0.1, affine=True, track_running_stats=True)
  (dropout1): Dropout(p=0.2, inplace=False)
  (dropout2): Dropout(p=0.35, inplace=False)
  (dropout3): Dropout(p=0.35, inplace=False)
  (linear1): Linear(in_features=150, out_features=96, bias=True)
  (linear2): Linear(in_features=96, out_features=96, bias=True)
  (linear3): Linear(in_features=96, out_features=48, bias=True)
  (linear4): Linear(in_features=48, out_features=2, bias=True)
  (leaky): LeakyReLU(negative_slope=0.01)
  (context_cnn): Sequential(
    (0): BatchNorm1d(150, eps=1e-05, momentum=0.1, affine=True, track_running_stats=True)
    (1): Conv1d(150, 96, kernel_size=(3,), stride=(1,), padding=(1,))
    (2): LeakyReLU(negative_slope=0.01)
    (3): Dropout(p=0.2, inplace=False)
    (4): BatchNorm1d(96, eps=1e-05, momentum=0.1, affine=True, track_running_stats=True)
    (5): Conv1d(96, 96, kernel_size=(3,), stride=(1,), padding=(1,))
    (6): LeakyReLU(negative_slope=0.01)
  )
  (convT): ConvTranspose1d(96, 2, kernel_size=(5,), stride=(1,))
  (distance_mlp): Sequential(
    (0): BatchNorm1d(150, eps=1e-05, momentum=0.1, affine=True, track_running_stats=True)
    (1): Linear(in_features=150, out_features=96, bias=True)
    (2): LeakyReLU(negative_slope=0.01)
    (3): BatchNorm1d(96, eps=1e-05, momentum=0.1, affine=True, track_running_stats=True)
    (4): Linear(in_features=96, out_features=96, bias=True)
    (5): LeakyReLU(negative_slope=0.01)
    (6): BatchNorm1d(96, eps=1e-05, momentum=0.1, affine=True, track_running_stats=True)
    (7): Linear(in_features=96, out_features=1, bias=True)
    (8): ReLU()
  )
)
\end{lstlisting}

\newpage
\section{Evaluation Results}
\label{sec:appendix:evaluation-results}

\begin{figure}[H]
    \centering
    \includegraphics[scale=0.5]{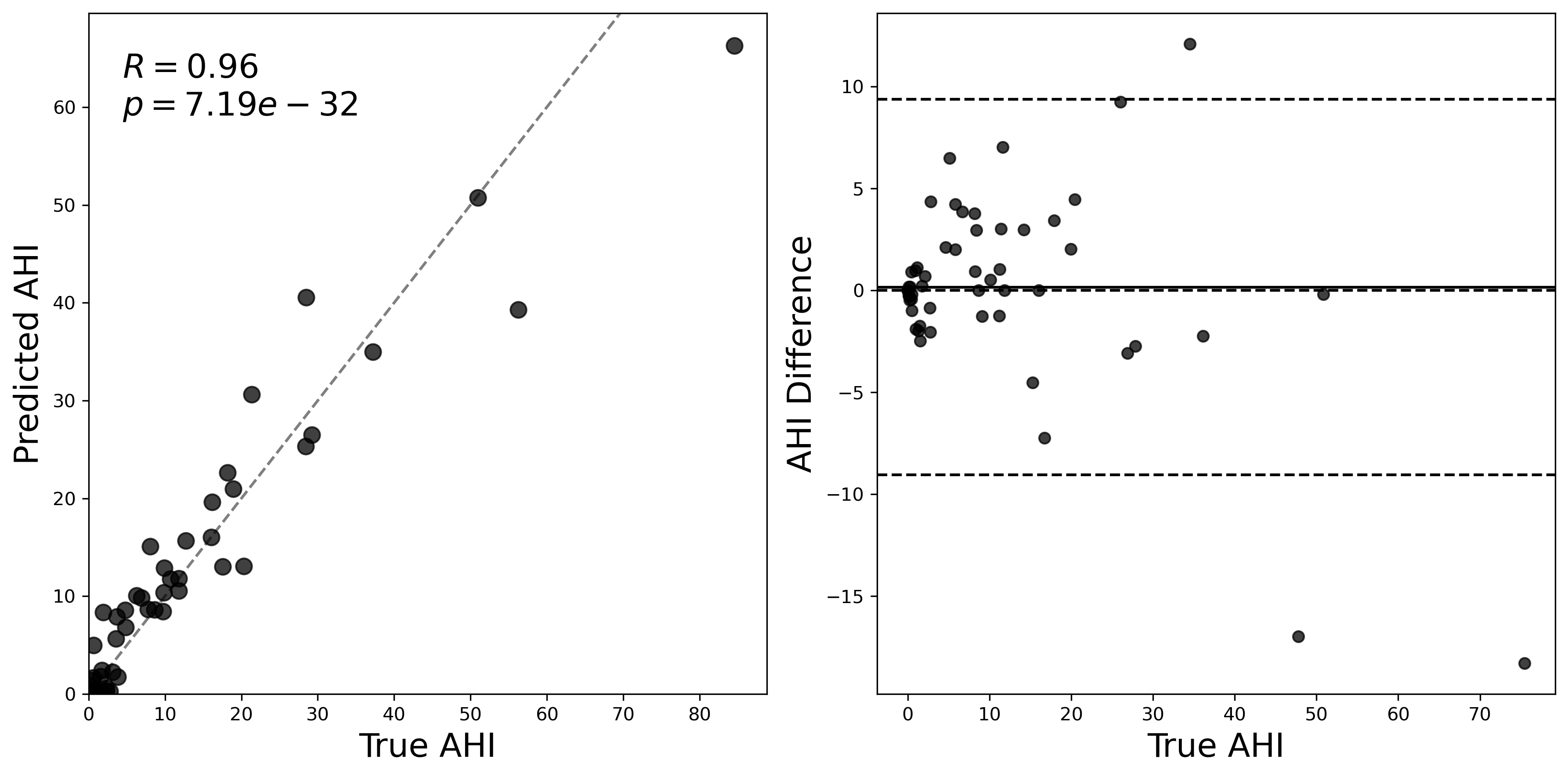}
    \vspace{-2mm}
    \caption{Scatterplot (left) and Bland-Altman plot (right) comparing the predicted and ground truth AHI across the test set when using the sleep intervals from the PSG annotation. For the Bland-Altman plot, the solid line is the bias and the dotted lines in the Bland-Altman plot represent 0 as well as 1.96 standard deviations above and below the mean.}
    \label{fig:sleep-scatter-blandaltman}
\end{figure}

\begin{figure}[H]
    \centering
    \includegraphics[scale=0.5]{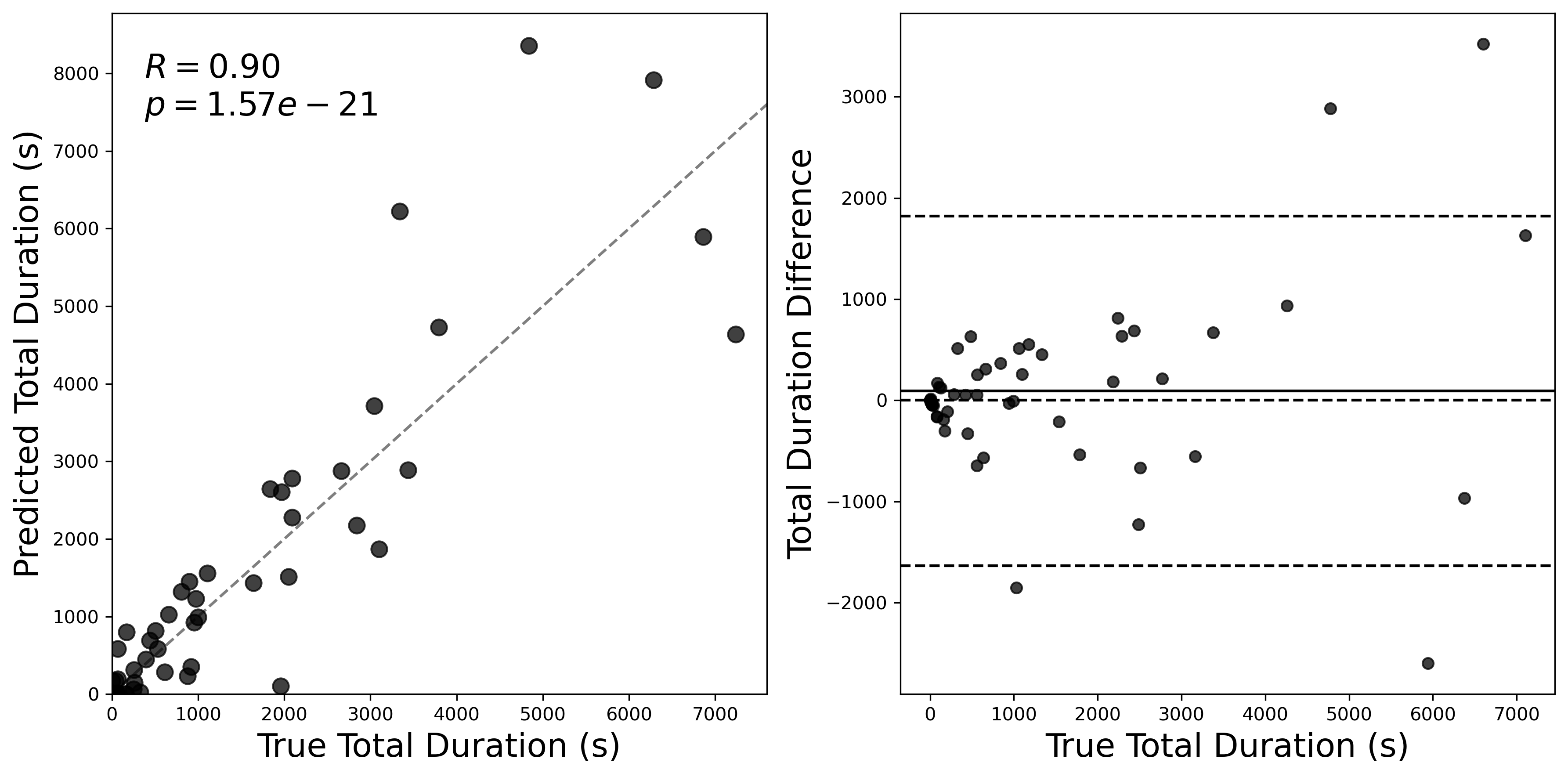}
    \vspace{-4mm}
    \caption{Scatterplot (left) and Bland-Altman plot (right) comparing the predicted and ground truth total apnea duration across the test set. For the Bland-Altman plot, the solid line is the bias and the dotted lines in the Bland-Altman plot represent 0 as well as 1.96 standard deviations above and below the mean.}
    \label{fig:total_apnea_duration}
\end{figure}


\begin{table}[H]
    \caption{Event-wise evaluation of the proposed model under punitive and non-punitive conditions with regards to over- and under-segmentation with varying IOU thresholds.}
    \vspace{-2mm}
    \resizebox{1\textwidth}{!}{
    \begin{tabular}{c|c|ccccc}
        \hline
         \textit{IOU Threshold} & \textit{Method} & \textit{F1} & \textit{Recall/Sens.} & \textit{Precision} & \textit{OSI * 100} & \textit{USI * 100} \\
        \hline
        \multirow{2}{*}{0.01} & \phantom{\Large{A}}Punitive\phantom{\Large{A}} & \makecell{70.00} & \makecell{72.68} & \makecell{67.52} & \makecell{1.40} & \makecell{3.36} \\
        & Non-punitive & \makecell{72.55} & \makecell{76.03} & \makecell{69.38} & \makecell{1.40} & \makecell{3.36} \\
        \hline
        
        \multirow{2}{*}{0.1} & \phantom{\Large{A}}Punitive\phantom{\Large{A}} & \makecell{69.76} & \makecell{72.42} & \makecell{67.28} & \makecell{1.22} & \makecell{3.17} \\
        & Non-punitive & \makecell{72.13} & \makecell{75.59} & \makecell{68.98} & \makecell{1.22} & \makecell{3.17} \\
        \hline

        \multirow{2}{*}{0.2} & \phantom{\Large{A}}Punitive\phantom{\Large{A}} & \makecell{69.33} & \makecell{71.98} & \makecell{66.87} & \makecell{0.885} & \makecell{2.69} \\
        & Non-punitive & \makecell{71.30} & \makecell{74.66} & \makecell{68.23} & \makecell{0.885} & \makecell{2.69} \\
        \hline

        \multirow{2}{*}{0.3} & \phantom{\Large{A}}Punitive\phantom{\Large{A}} & \makecell{67.91} & \makecell{70.50} & \makecell{65.50} & \makecell{0.295} & \makecell{1.55} \\
        & Non-punitive & \makecell{68.98} & \makecell{72.05} & \makecell{66.17} & \makecell{0.295} & \makecell{1.55} \\
        \hline

        \multirow{2}{*}{0.4} & \phantom{\Large{A}}Punitive\phantom{\Large{A}} & \makecell{64.71} & \makecell{67.18} & \makecell{62.42} & \makecell{0.111} & \makecell{0.258} \\
        & Non-punitive & \makecell{64.92} & \makecell{67.44} & \makecell{62.57} & \makecell{0.111} & \makecell{0.258} \\
        \hline

        \multirow{2}{*}{0.5} & \phantom{\Large{A}}Punitive\phantom{\Large{A}} & \makecell{59.60} & \makecell{61.87} & \makecell{57.49} & \makecell{0.00} & \makecell{0.00} \\
        & Non-punitive & \makecell{59.59} & \makecell{61.87} & \makecell{57.46} & \makecell{0.00} & \makecell{0.00} \\
        \hline
    \end{tabular}
    }
    \label{tab:iou_comparison}
\end{table}

\begin{figure}[H]
    \centering
    \includegraphics[scale=0.7]{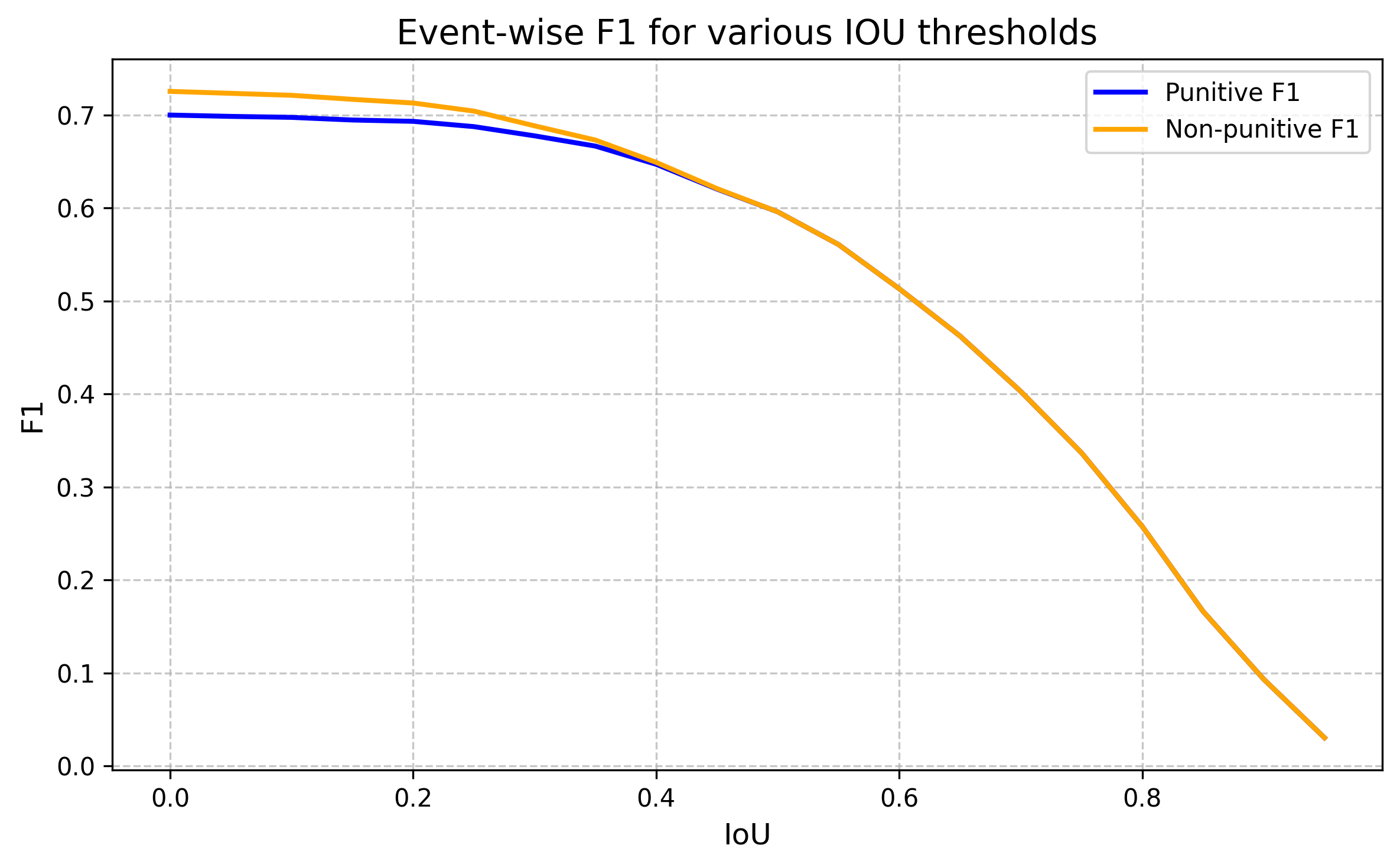}
    \vspace{-2mm}
    \caption{Effect of varying the event-wise IoU threshold on event-wise F1-score for both punitive and non-punitive approaches.}
    \label{fig:f1-vs-iou}
\end{figure}

\newpage
\section{More on Event-wise Metrics}
\label{sec:appendix:eventwise_metrics}

\subsection{Summary and Discussion of IoU Thresholds}

As a novel framework for the evaluation of event-detection models, the event-wise metrics warrant a more detailed explanation than the overview provided in Section \ref{sec:statistical_analysis}. To summarize, three new classes of metrics are introduced:
\begin{enumerate}
    \item Punitive event-wise metrics: Each true event is matched to one predicted event (if there is an overlapping one).
    \item Non-punitive event-wise metrics: Each true event is considered detected if any prediction sufficiently overlaps it.
    \item OSI and USI: Diagnostic indexes which measure how much a model fragments or merges true events when making predictions.
\end{enumerate}

We note that for each of these classes, a "sufficient overlap" criterion is required. Thus, we must define an IoU threshold for each class. For example, to determine whether a prediction and a true event are a match (a true positive), we check whether the true and predicted event share an IoU greater than our defined threshold. By choosing lower or higher IoU thresholds, we can adjust metrics to be less or more strict in what they consider a successful detection.\\

In this paper, the same IoU threshold is used across all three types of metrics. This is primarily for simplicity and consistency, and provides useful information due to the low threshold of 0.2, but we can imagine scenarios where nuances in threshold selection are significant. For example, we may want a high threshold for our punitive metrics to determine how many events we truly capture effectively, but we may want a low threshold for the non-punitive metrics and OSI/USI, since we suspect that our model also makes many small predictions that have small overlaps with true events. By keeping the criteria for a match low, the OSI and USI metrics may be much more informative, where they would otherwise ignore the many small overlapping fragments.\\

This example also leads us to a mathematical property of the framework, which is that for any IoU of greater than 0.5, the punitive and non-punitive metrics are equal, and the OSI and USI are 0. This is intuitive, since a 0.5 IoU threshold would allow no more than one event to be considered a successful overlap with another (there cannot be two predictions that each have 0.6 IoU with a true event). With this in mind, some recommendations can be established for future use of this event-wise metric framework, specifically in the types of IoU thresholds that will allow a reasonable or useful analysis:

\begin{enumerate}
    \item Punitive metrics may be used with any IoU threshold, depending on desired strictness of evaluation.
    \item Non-punitive metrics should only be used with IoU thresholds $<$ 0.5, ideally smaller values, to provide accurate insights into small fragments or merges.
    \item OSI/USI should be used with low IoU thresholds like 0-0.2, and kept constant even if thresholds for the punitive and non-punitive metrics are adjusted. This provides a consistent and objective insight into the model's over- and under-segmentation behavior.
\end{enumerate}

Note that it is not a requirement for the IoU thresholds of the non-punitive metrics and OSI/USI to be equal, as they exist for different purposes. It is also interesting to reiterate here the relation to TAO (True Apnea Overlap), which is equivalent to the recall of the non-punitive model with an IoU threshold of 0. Figure \ref{fig:f1-vs-iou} and Table \ref{tab:iou_comparison} show the effects of varying IoU on the metrics obtained for the present study's model.\\

Setting a higher IoU like 0.8-0.9 enforces stricter matching, meaning that the recall corresponds to the percentage of predictions that almost perfectly overlap with true events. By using a lower IoU of 0.20, we enforce matches less strictly. Thus, our main result in Table \ref{tab:EventWiseEval} should be interpreted as measuring whether the model predicts events at the generally correct times, as opposed to measuring perfect event durations, starts, or stops. As leading apnea-detection models improve, the IoU threshold may be increased to hold models to a higher standard for event-wise evaluation.\\

Although useful, this evaluation framework can benefit from more rigorous guidelines for IoU Selection, potentially even the necessary calculation of a plots like Figure \ref{fig:f1-vs-iou}. Furthermore, there are some edge cases, especially when there are many shorter predictions and true events in a short time, where the interpretability of these metrics is vague. \\

Nonetheless, for the present study, this framework provides a new and effective way to evaluate models on not only their segment-wise classification abilities or their AHI-predictivity, but on their ability to accurately detect individual sleep apnea events.

\subsection{Implementation Details}

Evaluating model performance on a per-event basis requires converting all model outputs and ground truth labels (which are per-second) to event intervals with start and end times (Algorithm \ref{alg:to-intervals}). Once this is done, algorithms \ref{alg:event-punitive} and \ref{alg:event-nonpun} are used to compute event-wise metrics that are punitive to fragmentation and non-punitive to fragmentation. These algorithms yield three basic metrics: the number of false positives, true positives, and false negatives. Lastly, the over-segmentation index and under-segmentation index (OSI and USI) can be computed with Algorithm \ref{alg:osi-usi}. Recall that although they are not direct performance metrics, OSI and USI can be useful in understanding where models make mistakes pertaining to fragmenting and merging predictions.

\begin{algorithm}
    \caption{Extract contiguous events from a binary sequence}
    \label{alg:to-intervals}
    \textbf{Input:} Binary labels $\mathbf{y}\in\{0,1\}^{T}$ \\
    \textbf{Output:} Event list $\mathcal{I}(\mathbf{y})=\{[s_k,e_k)\}_{k=1}^{K}$ of maximal runs of ones
    \begin{algorithmic}[1]
        \State $\mathcal{I} \gets [\,]$; \quad $k \gets 1$
        \While{$k \le T$}
            \If{$y_k = 1$}
                \State $s \gets k$
                \While{$k \le T$ \textbf{ and } $y_k = 1$} \State $k \gets k+1$ \EndWhile
                \State $e \gets k$ \Comment{Half-open $[s,e)$}
                \State append $[s,e)$ to $\mathcal{I}$
            \Else
                \State $k \gets k+1$
            \EndIf
        \EndWhile
        \State \textbf{return} $\mathcal{I}$
    \end{algorithmic}
\end{algorithm}

Algorithm \ref{alg:event-punitive} below matches predicted events to true events using the Hungarian matching algorithm with IoU(true, predicted) as a matching criterion. It performs no additional steps to account for multiple predicted events matching one true event or vice versa, and thus matches each true event to only one predicted event. In this sense, it is "punitive" to over- and under-segmentation.\\

\begin{algorithm}
    \caption{Event-wise metrics (Punitive one-to-one matching)}
    \label{alg:event-punitive}
    \textbf{Input:} True binaries $\{\mathbf{y}^{(n)}\}_{n=1}^{N}$, predicted binaries $\{\hat{\mathbf{y}}^{(n)}\}_{n=1}^{N}$, IoU threshold $\tau \in [0,1]$ \\
    \textbf{Output:} $TP,FP,FN$
    \begin{algorithmic}[1]
        \State $TP\gets 0,\; FP\gets 0,\; FN\gets 0$
        \For{$n=1$ \textbf{to} $N$}
            \State $\mathcal{I}_T \gets \mathcal{I}(\mathbf{y}^{(n)})$; \quad $\mathcal{I}_P \gets \mathcal{I}(\hat{\mathbf{y}}^{(n)})$ \Comment{Alg.~\ref{alg:to-intervals}}
            \State Build matrix $M\in\mathbb{R}^{|\mathcal{I}_T|\times|\mathcal{I}_P|}$ where $M_{ij} \gets \mathrm{IoU}(\mathcal{I}_T[i],\mathcal{I}_P[j])$
            \State $(\mathcal{A}_T,\mathcal{A}_P) \gets \textsc{Hungarian}(-M)$ \Comment{max IoU $\Leftrightarrow$ min cost $-M$}
            \State $\text{matched} \gets \{(i,j)\in \mathcal{A}_T\times\mathcal{A}_P \;|\; M_{ij}\ge \tau\}$
            \State $T_{\mathrm{cov}} \gets \{\, i \;|\; \exists j:(i,j)\in \text{matched}\,\}$; \quad $P_{\mathrm{cov}} \gets \{\, j \;|\; \exists i:(i,j)\in \text{matched}\,\}$
            \State $TP \gets TP + |T_{\mathrm{cov}}|$
            \State $FN \gets FN + (|\mathcal{I}_T| - |T_{\mathrm{cov}}|)$
            \State $FP \gets FP + (|\mathcal{I}_P| - |P_{\mathrm{cov}}|)$
        \EndFor
        \State \textbf{return} $TP,FP,FN$
    \end{algorithmic}
\end{algorithm}

Algorithm \ref{alg:event-nonpun} is non-punitive with respect to over- and under-segmentation. Instead of matching events one-to-one, it simply checks for sufficient overlap between true and predicted events. For example, one long predicted event spanning multiple true events can count as multiple true positives, as long as each true event shares some minimum IOU threshold with the prediction. An example of the difference between punitive and non-punitive event-wise metrics was provided in Figure \ref{fig:eventwise_example}.

\begin{algorithm}
    \caption{Event-wise metrics (Non-punitive coverage)}
    \label{alg:event-nonpun}
    \textbf{Input:} $\{\mathbf{y}^{(n)}\}_{n=1}^{N}$, $\{\hat{\mathbf{y}}^{(n)}\}_{n=1}^{N}$, IoU threshold $\tau$ \\
    \textbf{Output:} $TP^\star,FP^\star,FN^\star$
    \begin{algorithmic}[1]
        \State $TP^\star\gets 0,\; FP^\star\gets 0,\; FN^\star\gets 0$
        \For{$n=1$ \textbf{to} $N$}
            \State $\mathcal{I}_T \gets \mathcal{I}(\mathbf{y}^{(n)})$; \quad $\mathcal{I}_P \gets \mathcal{I}(\hat{\mathbf{y}}^{(n)})$
            \For{\textbf{each} $i \in \mathcal{I}_T$}
                \If{$\exists j\in \mathcal{I}_P \text{ s.t. } \mathrm{IoU}(i,j)\ge \tau$}
                    \State $TP^\star \gets TP^\star + 1$
                \Else
                    \State $FN^\star \gets FN^\star + 1$
                \EndIf
            \EndFor
            \For{\textbf{each} $j \in \mathcal{I}_P$}
                \If{$\nexists i\in \mathcal{I}_T \text{ s.t. } \mathrm{IoU}(i,j)\ge \tau$}
                    \State $FP^\star \gets FP^\star + 1$
                \EndIf
            \EndFor
        \EndFor
        \State \textbf{return} $TP^\star,FP^\star,FN^\star$
    \end{algorithmic}
\end{algorithm}

\begin{algorithm}
    \caption{Over-/Under-Segmentation (OSI/USI)}
    \label{alg:osi-usi}
    \textbf{Input:} $\mathcal{I}_T,\mathcal{I}_P$ for a recording; IoU threshold $\tau_s$ \\
    \textbf{Output:} $OSI,USI$
    \begin{algorithmic}[1]
        \State Build $M_{ij} \gets \mathrm{IoU}(\mathcal{I}_T[i],\mathcal{I}_P[j])$; \quad $(\mathcal{A}_T,\mathcal{A}_P) \gets \textsc{Hungarian}(-M)$
        \State $\text{matched} \gets \{(i,j)\in \mathcal{A}_T\times\mathcal{A}_P \;|\; M_{ij}\ge \tau_s\}$
        \State $E \gets 0$; \quad $S \gets 0$ \Comment{$E$ = extra predicted fragments; $S$ = swallowed trues}
        \For{\textbf{each} $(i,j)\in \text{matched}$}
            \State $E \gets E + \left|\{\, j'\neq j : \mathrm{IoU}(\mathcal{I}_T[i],\mathcal{I}_P[j']) \ge \tau_s \,\}\right|$
            \State $S \gets S + \left|\{\, i'\neq i : \mathrm{IoU}(\mathcal{I}_T[i'],\mathcal{I}_P[j]) \ge \tau_s \,\}\right|$
        \EndFor
        \State $OSI \gets \dfrac{E}{\max(|\mathcal{I}_T|,1)}$; \quad $USI \gets \dfrac{S}{\max(|\mathcal{I}_T|,1)}$
        \State \textbf{return} $OSI,USI$
    \end{algorithmic}
\end{algorithm}

\end{document}